\documentclass[12pt,twoside, a4paper]{article}
\def\pd{\partial}
\def\mc{\mathcal}

\usepackage[dvips]{graphicx}
\usepackage{amssymb}
\usepackage{amssymb,amsmath}
\usepackage{graphicx}
\usepackage{dsfont}
\usepackage{caption}
\usepackage{subcaption}
\input{epsf.sty}  
\begin{document}
\begin{center}
\LARGE{\textbf{Holographic RG flows in $N=4$ SCFTs from half-maximal gauged supergravity}}
\end{center}
\vspace{1 cm}
\begin{center}
\large{\textbf{Parinya Karndumri}$^a$ and \textbf{Khem
Upathambhakul}$^b$}
\end{center}
\begin{center}
String Theory and Supergravity Group, Department
of Physics, Faculty of Science, Chulalongkorn University, 254 Phayathai Road, Pathumwan, Bangkok 10330, Thailand
\end{center}
E-mail: $^a$parinya.ka@hotmail.com \\
E-mail: $^b$keima.tham@gmail.com \vspace{1 cm}\\
\begin{abstract}
We study four-dimensional $N=4$ gauged supergravity coupled to six vector multiplets with semisimple gauge groups $SO(4)\times SO(4)$, $SO(3,1)\times SO(3,1)$ and $SO(4)\times SO(3,1)$. All of these gauge groups are dyonically embedded in the global symmetry group $SO(6,6)$ via its maximal subgroup $SO(3,3)\times SO(3,3)$. For $SO(4)\times SO(4)$ gauge group, there are four $N=4$ supersymmetric $AdS_4$ vacua with $SO(4)\times SO(4)$, $SO(4)\times SO(3)$, $SO(3)\times SO(4)$ and $SO(3)\times SO(3)$ symmetries, respectively. These $AdS_4$ vacua correspond to $N=4$ SCFTs in three dimensions with $SO(4)$ R-symmetry and different flavor symmetries. We explicitly compute the full scalar mass spectra at all these vacua. Holographic RG flows interpolating between these conformal fixed points are also given. The solutions describe supersymmetric deformations of $N=4$ SCFTs by relevant operators of dimensions $\Delta=1,2$. A number of these solutions can be found analytically although some of them can only be obtained numerically. These results provide a rich and novel class of $N=4$ fixed points in three-dimensional Chern-Simons-Matter theories and possible RG flows between them in the framework of $N=4$ gauged supergravity in four dimensions. Similar studies are carried out for non-compact gauge groups, but the $SO(4)\times SO(4)$ gauge group exhibits a much richer structure.     
\end{abstract}
\newpage
\section{Introduction}
The study of holographic RG flows is one of the most interesting results from the celebrated AdS/CFT correspondence since its original proposal in \cite{maldacena}. The solutions take the form of domain walls interpolating between $AdS$ vacua, for RG flows between two conformal fixed points, or between an $AdS$ vacuum and a singular geometry, for RG flows from a conformal fixed point to a non-conformal field theory. Many of these fixed points are described by superconformal field theories (SCFTs) which are also believed to give some insight into the dynamics of various branes in string/M-theory.
\\
\indent Rather than finding holographic RG flow solutions directly in string/M-theory, a convenient and effective way to find these solutions is to look for domain wall solutions in lower dimensional gauged supergravities. In some cases, the resulting solutions can be uplifted to interesting brane configurations within string/M-theory, see for example \cite{Warner_membrane_flow,Warner_M_F_theory_flow,Warner_higher_Dflow}. Apart from rendering the computation simpler, working in lower dimensional gauged supergravities also has an advantage of being independent of higher dimensional embedding. Results obtained within this framework are applicable in any models described by the same effective gauged supergravity regardless of their higher dimensional origins.          
\\
\indent Many results along this direction have been found in maximally gauged supergravities, see for examples \cite{GPPZ,FGPP,Flow_in_N8_4D,4D_G2_flow,Warner_M2_flow,Warner_Fisch,Guarino_BPS_DW,Elec_mag_flows,Yi_4D_flow}. A number of RG flow solutions in half-maximally gauged supergravities in various dimensions have, on the other hand, been studied only recently in \cite{7Dflow,6Dflow,6Dflow1,5Dflow_Davide,5Dflow_bobev,tri-sasakian-flow,orbifold_flow,3Dflow}, see also \cite{Carlos_6D_flow} and \cite{Warner_5DN4_flow} for earlier results. In this paper, we will give holographic RG flow solutions within $N=4$ gauged supergravity in four dimensions. Solutions in the case of non-semisimple gauge groups with known higher dimensional origins have already been considered in \cite{tri-sasakian-flow} and \cite{orbifold_flow}. This non-semisimple gauging, however, turns out to have a very restricted number of supersymmetric $AdS_4$ vacua. In this work, we will consider semisimple gauging of $N=4$ supergravity similar to the study in other dimensions. Although some general properties of $AdS_4$ vacua and RG flows have been pointed out recently in \cite{5Dflow_bobev}, to the best of our knowledge, a detailed analysis and explicit RG flow solutions in $N=4$ gauged supergravity have not previously appeared.     
\\
\indent Gaugings of $N=4$ supergravity coupled to an arbitrary number $n$ of vector multiplets have been studied and classified for a long time \cite{Eric_N4_4D,de_Roo_N4_4D,N4_Wagemans}, and the embedding tensor formalism which includes all possible deformations of $N=4$ supergravity has been given in \cite{N4_gauged_SUGRA}. For the case of $n<6$, the relation between the resulting $N=4$ supergravity and ten-dimensional supergravity is not known. Therefore, we will consider only the case of $n\geq 6$ which is capable of embedding in ten dimensions. Furthermore, we are particularly intested in $N=4$ gauged supergravity coupled to six vector multiplets to simplify the computation. In this case, possible gauge groups are embedded in the global symmetry group $SL(2,\mathbb{R})\times SO(6,6)$. From a general result of \cite{AdS4_N4_Jan}, the existence of $N=4$ supersymmetric $AdS_4$ vacua requires that the gauge group is purely embedded in the $SO(6,6)$ factor. Furthermore, the gauge group must contain an $SO(3)\times SO(3)$ subgroup with one of the $SO(3)$ factors embedded electrically and the other one embedded magnetically. 
\\
\indent We will consider gauge groups in the form of a simple product $G_1\times G_2$ in which one of the two factors is embedded electrically in $SO(3,3)\subset SO(6,6)$ while the other is embedded magnetically in the other $SO(3,3)$ subgroup of $SO(6,6)$. Taking the above criterions for having supersymmetric $AdS_4$ vacua into account, we will study the case of $G_1,G_2=SO(4)$ and $SO(3,1)$. There are then three different product gauge groups to be considered, $SO(4)\times SO(4)$, $SO(3,1)\times SO(3,1)$ and $SO(4)\times SO(3,1)$. We will identify possible supersymmetric $AdS_4$ vacua and supersymmetric RG flows interpolating between these vacua. These solutions should describe RG flows in the dual $N=4$ Chern-Simons-Matter (CSM) theories driven by relevant operators dual to the scalar fields of the $N=4$ gauged supergravity. As shown in \cite{N4_CSM_Klebanov}, some of the $N=4$ CSM theories can be obtained from a non-chiral orbifold of the ABJM theory \cite{ABJM}. Other classes of $N=4$ CSM theories are also known, see \cite{N4_CSM_GW} and \cite{N4_CSM_Park} for example. These theories play an important role in describing the dynamics of M$2$-branes on various backgrounds. The solutions obtained in this paper should also be useful in this context via the AdS/CFT correspondence. It should also be  emphasized that all gauge groups considered here are currently not known to have higher dimensional origins. Therefore, the corresponding holographic duality in this case is still not firmly established.         
\\
\indent The paper is organized as follow. In section \ref{N4_SUGRA},
we review $N=4$ gauged supergravity coupled to vector multiplets in the embedding tensor formalism. This sets up the framework we will use throughout the paper and collects relevant formulae and notations used in subsequent sections. In section \ref{SO4_SO4_SUGRA}, $N=4$ gauged supergravity with $SO(4)\times SO(4)$ gauge group is constructed, and the scalar potential for scalars which are singlets under $SO(4)_{\textrm{inv}}\subset SO(4)\times SO(4)$ is computed. We will identify possible supersymmetric $AdS_4$ vacua and compute the full scalar mass spectra at these vacua. The section ends with supersymmetric RG flow solutions interpolating between $AdS_4$ vacua and RG flows to non-conformal field theories. A similar study is performed in sections \ref{SO3_1_SO3_1} and \ref{SO4_SO3_1} for non-compact $SO(3,1)\times SO(3,1)$ and $SO(4)\times SO(3,1)$ gauge groups. Conclusions and comments on the results will be given in section \ref{conclusion}. An appendix containing the convention on `t Hooft matrices is included at the end of the paper.

\section{$N=4$ gauged supergravity coupled to vector multiplets}\label{N4_SUGRA} 
To set up our framework, we give a brief review of four-dimensional $N=4$ gauged
supergravity. We mainly give relevant information and necessary formulae to find supersymmetric $AdS_4$ vacua and domain wall solutions. More details on the construction can be found in
\cite{N4_gauged_SUGRA}. 
\\
\indent $N=4$ supergravity can couple to an arbitrary number $n$ of vector multiplets. The supergravity multiplet consists of the graviton
$e^{\hat{\mu}}_\mu$, four gravitini $\psi^i_\mu$, six vectors
$A_\mu^m$, four spin-$\frac{1}{2}$ fields $\chi^i$ and one complex
scalar $\tau$ containing the dilaton $\phi$ and the axion $\chi$. The complex scalar can be parametrized by $SL(2,\mathbb{R})/SO(2)$ coset. Each vector multiplet contains a vector field $A_\mu$, four gaugini $\lambda^i$ and six scalars $\phi^m$.
Similar to the dilaton and the axion in the gravity multiplet, the $6n$ scalar fields can be parametrized by $SO(6,n)/SO(6)\times SO(n)$ coset.
\\
\indent Throughout the paper, space-time and tangent space indices are denoted respectively by $\mu,\nu,\ldots =0,1,2,3$ and
$\hat{\mu},\hat{\nu},\ldots=0,1,2,3$. The $SO(6)\sim SU(4)$
R-symmetry indices will be described by $m,n=1,\ldots, 6$ for the
$SO(6)$ vector representation and $i,j=1,2,3,4$ for the $SO(6)$
spinor or $SU(4)$ fundamental representation. The $n$ vector
multiplets will be labeled by indices $a,b=1,\ldots, n$. All fields in the vector multiplets accordingly carry an additional
index in the form of $(A^a_\mu,\lambda^{ia},\phi^{ma})$. 
\\
\indent Fermionic fields and the supersymmetry parameters transforming in the fundamental representation of $SU(4)_R\sim SO(6)_R$ R-symmetry are subject to the chirality projections
\begin{equation}
\gamma_5\psi^i_\mu=\psi^i_\mu,\qquad \gamma_5\chi^i=-\chi^i,\qquad \gamma_5\lambda^i=\lambda^i
\end{equation}
while the fields transforming in the anti-fundamental representation of $SU(4)_R$ satisfy
\begin{equation}
\gamma_5\psi_{\mu i}=-\psi_{\mu i},\qquad \gamma_5\chi_i=\chi_i,\qquad \gamma_5\lambda_i=-\lambda_i\, .
\end{equation}
\indent Gaugings of the matter-coupled $N=4$ supergravity can be described by two components of the embedding tensor $\xi^{\alpha M}$ and $f_{\alpha MNP}$ with $\alpha=(+,-)$ and $M,N=(m,a)=1,\ldots, n+6$ denoting fundamental
representations of $SL(2,\mathbb{R})$ and $SO(6,n)$, respectively. Under the full $SL(2,\mathbb{R})\times SO(6,n)$ duality symmetry, the electric vector fields $A^{+M}=(A^m_\mu,A^a_\mu)$, appearing in the ungauged Lagrangian, together with their magnetic dual $A^{-M}$ form a doublet under $SL(2,\mathbb{R})$ denoted by $A^{\alpha M}$. A general gauge group is embedded in both $SL(2,\mathbb{R})$ and
$SO(6,n)$, and the magnetic vector fields can also participate in the gauging. However, each magnetic vector field must be accompanied by an auxiliary two-form field in order to remove the extra degrees of freedom.
\\
\indent From the analysis of supersymmetric $AdS_4$ vacua in \cite{AdS4_N4_Jan}, see also \cite{de_Roo_N4_4D,N4_Wagemans}, purely electric gaugings do not admit $AdS_4$ vacua unless an $SL(2,\mathbb{R})$ phase is included \cite{de_Roo_N4_4D}. The latter is however incorporated in the magnetic component $f_{-MNP}$ \cite{N4_gauged_SUGRA}. Therefore, only gaugings involving both
electric and magnetic vector fields, or dyonic gaugings, lead to $AdS_4$ vacua. Furthermore, the existence of maximally supersymmetric $AdS_4$ vacua requires $\xi^{\alpha M}=0$. Accordingly, we will from now on restrict ourselves to the case of dyonic gaugings and $\xi^{\alpha M}=0$.
\\
\indent With $\xi^{\alpha M}=0$, the gauge covariant derivative can be written as
\begin{equation}
D_\mu=\nabla_\mu-gA_\mu^{\alpha M}f_{\alpha M}^{\phantom{\alpha M}NP}t_{NP}
\end{equation}
where $\nabla_\mu$ is the usual space-time covariant derivative including the spin connection.
$t_{MN}$ are $SO(6,n)$ generators in the fundamental representation and can be chosen as
\begin{equation}
(t_{MN})_P^{\phantom{P}Q}=2\delta^Q_{[M}\eta_{N]P}\, .
\end{equation}
$\eta_{MN}=\textrm{diag}(-1,-1,-1,-1,-1,-1,1,\ldots,1)$ is the $SO(6,n)$ invariant tensor, and $g$ is the gauge coupling constant
that can be absorbed in the embedding tensor $f_{\alpha MNP}$. For a product gauge group consisting of many simple subgroups, there can be as many independent coupling constants as the simple groups within the product. Note also that with the component $\xi^{\alpha M}=0$, the gauge group is embedded solely in $SO(6,n)$.
\\
\indent To define a consistent gauging, the embedding tensor has to satisfy a set of quadratic constraints
\begin{equation}
f_{\alpha R[MN}f_{\beta PQ]}^{\phantom{\beta PQ}R}=0,\qquad \epsilon^{\alpha\beta}f_{\alpha MNR}f_{\beta PQ}^{\phantom{\beta PQ}R}=0\, .
\end{equation}
as well as the linear or representation constraint $f_{\alpha MNP}=f_{\alpha[MNP]}$.
\\
\indent The scalar coset $SL(2,\mathbb{R})/SO(2)$ can be described by the coset representative $\mc{V}_\alpha$. We will choose the explicit form of $\mc{V}_\alpha$ as follow
\begin{equation}
\mc{V}_\alpha=e^{\frac{\phi}{2}}\left(
                                         \begin{array}{c}
                                           \chi-ie^{-\phi} \\
                                           1 \\
                                         \end{array}
                                       \right).
\end{equation}
\indent For the $SO(6,n)/SO(6)\times SO(n)$ coset, we introduce a coset representative $\mc{V}_M^{\phantom{M}A}$ transforming under global $SO(6,n)$ and local $SO(6)\times SO(n)$ by left and right multiplications, respectively. By splitting the index $A=(m,a)$, we can write the coset representative as
\begin{equation}
\mc{V}_M^{\phantom{M}A}=(\mc{V}_M^{\phantom{M}m},\mc{V}_M^{\phantom{M}a}).
\end{equation}
Being an element of $SO(6,n)$, the matrix $\mc{V}_M^{\phantom{M}A}$ satisfies the relation
\begin{equation}
\eta_{MN}=-\mc{V}_M^{\phantom{M}m}\mc{V}_N^{\phantom{M}m}+\mc{V}_M^{\phantom{M}a}\mc{V}_N^{\phantom{M}a}\,
.
\end{equation}
In addition, we can parametrize the $SO(6,n)/SO(6)\times SO(n)$ coset in term of a symmetric matrix
\begin{equation}
M_{MN}=\mc{V}_M^{\phantom{M}m}\mc{V}_N^{\phantom{M}m}+\mc{V}_M^{\phantom{M}a}\mc{V}_N^{\phantom{M}a}
\end{equation}
which is manifestly $SO(6)\times SO(n)$ invariant.
\\
\indent In this paper, we are mainly interested in supersymmetric solutions with only the metric and scalars non-vanishing. The bosonic Lagrangian with vector and auxiliary two-form fields vanishing can be written as
\begin{equation}
e^{-1}\mc{L}=\frac{1}{2}R+\frac{1}{16}\pd_\mu M_{MN}\pd^\mu
M^{MN}-\frac{1}{4(\textrm{Im}\tau)^2}\pd_\mu \tau \pd^\mu \tau^*-V
\end{equation}
where $e$ is the vielbein determinant. The scalar potential is given in terms of the scalar coset representative and the embedding tensor by
\begin{eqnarray}
V&=&\frac{g^2}{16}\left[f_{\alpha MNP}f_{\beta
QRS}M^{\alpha\beta}\left[\frac{1}{3}M^{MQ}M^{NR}M^{PS}+\left(\frac{2}{3}\eta^{MQ}
-M^{MQ}\right)\eta^{NR}\eta^{PS}\right]\right.\nonumber \\
& &\left.-\frac{4}{9}f_{\alpha MNP}f_{\beta
QRS}\epsilon^{\alpha\beta}M^{MNPQRS}\right]
\end{eqnarray}
where $M^{MN}$ is the inverse of $M_{MN}$, and $M^{MNPQRS}$ is obtained from
\begin{equation}
M_{MNPQRS}=\epsilon_{mnpqrs}\mc{V}_{M}^{\phantom{M}m}\mc{V}_{N}^{\phantom{M}n}
\mc{V}_{P}^{\phantom{M}p}\mc{V}_{Q}^{\phantom{M}q}\mc{V}_{R}^{\phantom{M}r}\mc{V}_{S}^{\phantom{M}s}\label{M_6}
\end{equation}
with indices raised by $\eta^{MN}$. Similar to $M^{MN}$, $M^{\alpha\beta}$ is the inverse of a symmetric $2\times 2$ matrix $M_{\alpha\beta}$ defined by
\begin{equation}
M_{\alpha\beta}=\textrm{Re}(\mc{V}_\alpha\mc{V}^*_\beta).
\end{equation}
\indent Fermionic supersymmetry transformations are given by
\begin{eqnarray}
\delta\psi^i_\mu &=&2D_\mu \epsilon^i-\frac{2}{3}gA^{ij}_1\gamma_\mu
\epsilon_j,\\
\delta \chi^i &=&i\epsilon^{\alpha\beta}\mc{V}_\alpha D_\mu
\mc{V}_\beta\gamma^\mu \epsilon^i-\frac{4}{3}igA_2^{ij}\epsilon_j,\\
\delta \lambda^i_a&=&2i\mc{V}_a^{\phantom{a}M}D_\mu
\mc{V}_M^{\phantom{M}ij}\gamma^\mu\epsilon_j+2igA_{2aj}^{\phantom{2aj}i}\epsilon^j\,
.
\end{eqnarray}
The fermion shift matrices, also appearing in fermionic mass-like terms of the gauged Lagrangian, are given by
\begin{eqnarray}
A_1^{ij}&=&\epsilon^{\alpha\beta}(\mc{V}_\alpha)^*\mc{V}_{kl}^{\phantom{kl}M}\mc{V}_N^{\phantom{N}ik}
\mc{V}_P^{\phantom{P}jl}f_{\beta M}^{\phantom{\beta M}NP},\nonumber
\\
A_2^{ij}&=&\epsilon^{\alpha\beta}\mc{V}_\alpha\mc{V}_{kl}^{\phantom{kl}M}\mc{V}_N^{\phantom{N}ik}
\mc{V}_P^{\phantom{P}jl}f_{\beta M}^{\phantom{\beta M}NP},\nonumber
\\
A_{2ai}^{\phantom{2ai}j}&=&\epsilon^{\alpha\beta}\mc{V}_\alpha
\mc{V}^M_{\phantom{M}a}\mc{V}^N_{\phantom{N}ik}\mc{V}_P^{\phantom{P}jk}f_{\beta
MN}^{\phantom{\beta MN}P}
\end{eqnarray}
where $\mc{V}_M^{\phantom{M}ij}$ is defined in terms of the 't Hooft
symbols $G^{ij}_m$ and $\mc{V}_M^{\phantom{M}m}$ as
\begin{equation}
\mc{V}_M^{\phantom{M}ij}=\frac{1}{2}\mc{V}_M^{\phantom{M}m}G^{ij}_m\, .
\end{equation}
Similarly, the inverse elements $\mc{V}^M_{\phantom{M}ij}$ can be written as
\begin{equation}
\mc{V}^M_{\phantom{M}ij}=-\frac{1}{2}\mc{V}^M_{\phantom{M}m}(G^{ij}_m)^*\,
.
\end{equation}
$G^{ij}_m$ convert an index $m$ in the vector representation of $SO(6)$ into an anti-symmetric pair of indices $[ij]$ in the $SU(4)$ fundamental representation. They satisfy the relations
\begin{equation}
G_{mij}=-(G^{ij}_m)^*=-\frac{1}{2}\epsilon_{ijkl}G^{kl}_m\, .
\end{equation}
The explicit form of these matrices can be found in the appendix.
\\
\indent The scalar potential can also be written in terms of the fermion shift matrices $A_1$ and $A_2$ as
\begin{equation}
V=-\frac{1}{3}A^{ij}_1A_{1ij}+\frac{1}{9}A^{ij}_2A_{2ij}+\frac{1}{2}A_{2ai}^{\phantom{2ai}j}
A_{2a\phantom{i}j}^{\phantom{2a}i}\, .
\end{equation}
Together with the fermionic supersymmetry transformations, it then follows that unbroken supersymmetry corresponds to an eigenvalue of $A^{ij}_1$, $\alpha$, satisfying $V_0=-\frac{\alpha^2}{3}$ where $V_0$ is the value of the scalar potential at the vacuum or the cosmological constant.

\section{Supersymmetric $AdS_4$ vacua and holographic RG flows in $SO(4)\times SO(4)$ gauged supergravity}\label{SO4_SO4_SUGRA}
We are interested in gauge groups that can be embedded in $SO(3,3)\times SO(3,3)\subset SO(6,6)$. These gauge groups take the form of a product $G_1\times G_2$ with $G_1,G_2\subset SO(3,3)$ being six-dimensional. Semisimple groups of dimension six that can be embedded in $SO(3,3)$ are $SO(4)$, $SO(3,1)$ and $SO(2,2)$. The embedding tensors for these gauge groups are given in \cite{dS_Roest}. Since gauge groups involving $SO(2,2)$ factors do not give rise to $AdS_4$ vacua, we will not consider these gauge groups in this paper. In this section, we will study $N=4$ gauged supergravity with compact $SO(4)\times SO(4)\sim SO(3)\times SO(3)\times SO(3)\times SO(3)$ gauge group. 

\subsection{$AdS_4$ vacua} 
Non-vanishing components of the embedding tensor for $SO(4)\times SO(4)$ gauge group are given by
\begin{eqnarray}
f_{+\hat{m}\hat{n}\hat{p}}&=&\sqrt{2}(g_1-\tilde{g}_1)\epsilon_{\hat{m}\hat{n}\hat{p}},\qquad f_{+\hat{a}\hat{b}\hat{c}}=\sqrt{2}(g_1+\tilde{g}_1)\epsilon_{\hat{a}\hat{b}\hat{c}},\nonumber \\
f_{-\tilde{m}\tilde{n}\tilde{p}}&=&\sqrt{2}(g_2-\tilde{g}_2)\epsilon_{\tilde{m}\tilde{n}\tilde{p}},\qquad f_{-\tilde{a}\tilde{b}\tilde{c}}=\sqrt{2}(g_2+\tilde{g}_2)\epsilon_{\tilde{a}\tilde{b}\tilde{c}},
\end{eqnarray}
where we have used the indices $M=(m,a)=(\hat{m},\tilde{m},\hat{a},\tilde{a})$ with $\hat{m}=1,2,3$, $\tilde{m}=4,5,6$, $\hat{a}=7,8,9$ and $\tilde{a}=10,11,12$. As mentioned before, the first and second $SO(4)$ factors are embedded electrically and magnetically, respectively. 
\\
\indent Non-compact generators of $SO(6,6)$ are given by
\begin{equation}
Y_{ma}=e_{m,a+6}+e_{a+6,m}
\end{equation}
in which the $12\times 12$ matrices $e_{MN}$ are defined by
\begin{equation}
(e_{MN})_{PQ}=\delta_{MP}\delta_{NQ}\, .
\end{equation}
The $36$ scalars in $SO(6,6)/SO(6)\times SO(6)$ transform as $(\mathbf{6},\mathbf{6})$ under the compact group $SO(6)\times SO(6)$.
Under the gauge group $SO(4)_+\times SO(4)_-\sim SO(3)^1_{+}\times SO(3)^2_{+}\times SO(3)^1_{-}\times SO(3)^2_{-}$, these scalars transform as
\begin{equation}
(\mathbf{6},\mathbf{6})\rightarrow (\mathbf{3},\mathbf{3},\mathbf{1},\mathbf{1})+(\mathbf{3},\mathbf{1},\mathbf{1},\mathbf{3})+(\mathbf{1},\mathbf{3},\mathbf{3},\mathbf{1})+(\mathbf{1},\mathbf{1},\mathbf{3},\mathbf{3}).
\end{equation}
\indent We will consider scalars which are singlets under the diagonal subgroup $SO(4)_{\textrm{inv}}\sim [SO(3)^1_{+}\times SO(3)^2_{+}]_D\times [SO(3)^1_{-}\times SO(3)^2_{-}]_D$. Under $SO(4)_{\textrm{inv}}$, the scalars transform as
\begin{equation}
2(\mathbf{1},\mathbf{1})+(\mathbf{3},\mathbf{1})+(\mathbf{1},\mathbf{3})+(\mathbf{1},\mathbf{5})+(\mathbf{5},\mathbf{1})+2(\mathbf{3},\mathbf{3}).
\end{equation}
The two singlets correspond to the following non-compact generators
\begin{equation}
\hat{Y}_1=Y_{11}+Y_{22}+Y_{33},\qquad \hat{Y}_2=Y_{44}+Y_{55}+Y_{66}\label{SO4_inv_L}
\end{equation}
in terms of which the coset representative is given by
\begin{equation}
L=e^{\phi_1\hat{Y}_1}e^{\phi_2\hat{Y}_2}\, .\label{L_SO4_inv}
\end{equation}
\indent Together with the $SL(2,\mathbb{R})/SO(2)$ scalars which are $SO(4)\times SO(4)$ singlets, there are four $SO(4)_{\textrm{inv}}$ singlet scalars. The scalar potential for these singlets can be computed to be
\begin{eqnarray}
V&=&\frac{1}{8}e^{-\phi-6\phi_1-6\phi_2}\left[e^{\phi+3\phi_2}\left[e^{\phi+3\phi_2}g_1^2+e^{\phi+12\phi_1+3\phi_2}\tilde{g}^2_1-3e^{\phi+4\phi_1
+3\phi_2}(2g_1^2+\tilde{g}^2_1)\right.\right. \nonumber \\
& &-3e^{\phi+8\phi_1+3\phi_2}(g_1^2+2\tilde{g}^2_1)+2e^{3\phi_1}g_1[g_2(1+3e^{4\phi_2})-e^{2\phi_2}\tilde{g}_2(3+e^{4\phi_2})]\nonumber \\
& &+6e^{7\phi_1}g_1[g_2(1+3e^{4\phi_2})-e^{2\phi_2}\tilde{g}_2(3+e^{4\phi_2})]-6e^{5\phi_1}\tilde{g}_1[g_2(1+3e^{4\phi_2})\nonumber \\
& &\left. -e^{2\phi_2}\tilde{g}_2(3+e^{4\phi_2})-2e^{9\phi_1}\tilde{g}_1[g_2(1+3e^{4\phi_2})-e^{2\phi_2}\tilde{g}_2(3+e^{4\phi_2})]\right]\nonumber \\
& &+e^{6\phi_1}\left[[1-3e^{4\phi_2}(2+e^{4\phi_2})](1+e^{2\phi}\chi^2)g_2^2+16g_2\tilde{g_2}e^{6\phi_2}(1+e^{2\phi}\chi^2)\right. \nonumber \\
& &+e^{4\phi_2}\left[e^{2\phi}(16e^{2\phi_2}g_1\tilde{g}_1-3\tilde{g}_2^2\chi^2-6\tilde{g}_2^2e^{4\phi_2}\chi^2+e^{8\phi_2}\chi^2\tilde{g}_2^2)\nonumber \right.\nonumber \\
& &\left.\left.\left. +(e^{8\phi_2}-6e^{4\phi_2}-3)\tilde{g}_2^2\right]\right]\right].\label{SO4_potential}
\end{eqnarray}
\indent This potential admits a maximally supersymmetric $AdS_4$ critical point with $SO(4)\times SO(4)$ symmetry at $\chi=\phi_1=\phi_2=0$ and 
\begin{equation}
\phi=\ln \left|\frac{g_2-\tilde{g}_2}{g_1-\tilde{g}_1}\right| .
\end{equation}
For convenience, we will denote this $AdS_4$ vacuum with $SO(4)\times SO(4)$ symmetry by critical point I.
\\
\indent Without loss of generality, we can shift the dilaton such that this critical point occurs at $\phi=0$. For definiteness, we will choose
\begin{equation}
\tilde{g}_2=g_1+g_2-\tilde{g}_1\, .
\end{equation}
At this critical point, we find the value of the cosmological constant and the $AdS_4$ radius
\begin{equation}
V_0=-6(g_1-\tilde{g}_1)^2\qquad\textrm{and}\qquad L=\frac{1}{\sqrt{2}(\tilde{g}_1-g_1)}
\end{equation}
where we have assumed that $\tilde{g}_1>g_1$. All scalars have masses $m^2L^2=-2$. In general, using the relation $m^2L^2=\Delta(\Delta-3)$, we find that these scalars can be dual to operators of dimensions $\Delta=1$ or $\Delta=2$. These correspond to mass terms of scalars ($\Delta =1$) or fermions ($\Delta=2$) in the dual three-dimensional SCFTs. Usually, the correct choice is fixed by supersymmetry as in the case of ABJM theory. However, in the present case, the identifcation is not so clear.  
\\
\indent Furthermore, the scalar potential in \eqref{SO4_potential} also admits additional three supersymmetric $AdS_4$ vacua:  
\begin{itemize}
\item II. This critical point has $SO(3)_+\times SO(4)_-$ symmetry with
\begin{eqnarray}
\phi&=&\ln\left[\frac{2\sqrt{g_1\tilde{g}_1}}{g_1+\tilde{g}_1}\right],\qquad \phi_1=\frac{1}{2}\ln\left[\frac{g_1}{\tilde{g}_1}\right],\qquad \phi_2=0,\nonumber \\
V_0&=&-\frac{3(g_1+\tilde{g}_1)(g_1-\tilde{g}_1)^2}{\sqrt{g_1\tilde{g}_1}},\qquad L=\frac{(g_1\tilde{g}_1)^{\frac{1}{4}}}{(\tilde{g}_1-g_1)\sqrt{g_1+\tilde{g}_1}}\, .
\end{eqnarray}
\item III. This critical point has $SO(4)_+\times SO(3)_-$ symmetry with
\begin{eqnarray}
\phi&=&-\ln\left[\frac{2\sqrt{g_2\tilde{g}_2}}{g_2+\tilde{g}_2}\right],\qquad \phi_2=\frac{1}{2}\ln\left[\frac{g_2}{\tilde{g}_2}\right],\qquad \phi_1=0,\nonumber \\
V_0&=&-\frac{3(g_2+\tilde{g}_2)(g_1-\tilde{g}_1)^2}{\sqrt{g_2\tilde{g}_2}},\qquad L=\frac{(g_2\tilde{g}_2)^{\frac{1}{4}}}{(\tilde{g}_1-g_1)\sqrt{g_2+\tilde{g}_2}}\, .
\end{eqnarray}
\item IV. This critical point is invariant under a smaller symmetry $SO(4)_{\textrm{inv}}$ with
\begin{eqnarray}
\phi&=&\ln\left[\sqrt{\frac{g_1\tilde{g}_1}{g_2\tilde{g}_2}}\frac{g_2+\tilde{g}_2}{g_1+\tilde{g}_1}\right],\qquad \phi_1=\frac{1}{2}\ln\left[\frac{g_1}{\tilde{g}_1}\right],\qquad \phi_2=\frac{1}{2}\ln\left[\frac{g_2}{\tilde{g}_2}\right],\nonumber \\
V_0&=&-\frac{3(g_2+\tilde{g}_2)^2(g_1-\tilde{g}_1)^2}{2\sqrt{g_1\tilde{g}_1g_2\tilde{g}_2}},\quad L=\frac{\sqrt{2}(g_1\tilde{g}_1g_2\tilde{g}_2)^{\frac{1}{4}}}{(\tilde{g}_1-g_1)\sqrt{(g_1+\tilde{g}_1)(g_2+\tilde{g}_2)}}\, .\quad 
\end{eqnarray}
\end{itemize}
We have written the above equations in term of $\tilde{g}_2$ for brevity.  All of these critical points preserve the full $N=4$ supersymmetry and correspond to $N=4$ SCFTs in three dimensions. Scalar masses at these critical points are given in tables \ref{table1}, \ref{table2} and \ref{table3}. 
\\
\indent It should also be noted that three massless scalars at critical points II and III are Goldstone bosons corresponding to the symmetry breaking $SO(4)_+\times SO(4)_-\rightarrow SO(3)_+\times SO(4)_-$ and $SO(4)_+\times SO(4)_-\rightarrow SO(4)_+\times SO(3)_-$. These scalars live in representations $(\mathbf{3},\mathbf{1},\mathbf{1})$ and $(\mathbf{1},\mathbf{1},\mathbf{3})$, respectively. Similarly, for critical point IV, six of the massless scalars in the representation $(\mathbf{1},\mathbf{3})+(\mathbf{3},\mathbf{1})$ are Goldstone bosons of the symmetry breaking $SO(4)_+\times SO(4)_-\rightarrow SO(4)_{\textrm{inv}}$. The remaining massless scalars correspond to marginal deformations in the SCFTs dual to these $AdS_4$ vacua. These deformations necessarily break some amount of supersymmetry since the $N=4$ $AdS_4$ vacua have no moduli preserving $N=4$ supersymmetry \cite{AdS4_N4_Jan}. It should also be noted that the vacuum structure of this gauged supergravity is very similar to two copies of $SO(3)\times SO(3)\sim SO(4)$ $N=3$ gauged supergravity considered in \cite{N3_4D_gauging}.
\\
\begin{table}[h]
\centering
\begin{tabular}{|c|c|c|}
  \hline
  Scalar field representations & $m^2L^2\phantom{\frac{1}{2}}$ & $\Delta$  \\ \hline
  $(\mathbf{1},\mathbf{1},\mathbf{1})$ & $-2_{\times 2}$ &  $1,2$  \\
  $(\mathbf{1},\mathbf{1},\mathbf{1})$ & $4$ &  $4$  \\
   $(\mathbf{3},\mathbf{1},\mathbf{1})$ & $0_{\times 3}$ &  $3$  \\
    $(\mathbf{1},\mathbf{3},\mathbf{3})$ & $0_{\times 9}$ &  $3$  \\
      $(\mathbf{5},\mathbf{1},\mathbf{1})$ & $-2_{\times 5}$ &  $1,2$  \\
   $(\mathbf{3},\mathbf{1},\mathbf{3})+(\mathbf{3},\mathbf{3},\mathbf{1})$ & $-2_{\times 18}$ &  $1,2$  \\
  \hline
\end{tabular}
\caption{Scalar masses at the $N=4$ supersymmetric $AdS_4$ critical
point with $SO(3)_+\times SO(4)_-$ symmetry and the
corresponding dimensions of the dual operators.}\label{table1}
\end{table}

\begin{table}[h]
\centering
\begin{tabular}{|c|c|c|}
  \hline
  Scalar field representations & $m^2L^2\phantom{\frac{1}{2}}$ & $\Delta$  \\ \hline
  $(\mathbf{1},\mathbf{1},\mathbf{1})$ & $-2_{\times 2}$ &  $1,2$  \\
  $(\mathbf{1},\mathbf{1},\mathbf{1})$ & $4$ &  $4$  \\
   $(\mathbf{1},\mathbf{1},\mathbf{3})$ & $0_{\times 3}$ &  $3$  \\
    $(\mathbf{3},\mathbf{3},\mathbf{1})$ & $0_{\times 9}$ &  $3$  \\
      $(\mathbf{1},\mathbf{1},\mathbf{5})$ & $-2_{\times 5}$ &  $1,2$  \\
   $(\mathbf{1},\mathbf{3},\mathbf{3})+(\mathbf{3},\mathbf{1},\mathbf{3})$ & $-2_{\times 18}$ &  $1,2$  \\
  \hline
\end{tabular}
\caption{Scalar masses at the $N=4$ supersymmetric $AdS_4$ critical
point with $SO(4)_+\times SO(3)_-$ symmetry and the
corresponding dimensions of the dual operators.}\label{table2}
\end{table}

\begin{table}[h]
\centering
\begin{tabular}{|c|c|c|}
  \hline
  Scalar field representations & $m^2L^2\phantom{\frac{1}{2}}$ & $\Delta$  \\ \hline
  $(\mathbf{1},\mathbf{1})$ & $-2_{\times 2}$ &  $1,2$  \\
  $(\mathbf{1},\mathbf{1})$ & $4_{\times 2}$ &  $4$  \\
   $(\mathbf{1},\mathbf{5})+(\mathbf{5},\mathbf{1})$ & $-2_{\times 10}$ &  $1,2$  \\
  $(\mathbf{1},\mathbf{3})+(\mathbf{3},\mathbf{1})$ & $0_{\times 6}$ &  $3$  \\
  $(\mathbf{3},\mathbf{3})$ & $0_{\times 18}$ &  $3$  \\
  \hline
\end{tabular}
\caption{Scalar masses at the $N=4$ supersymmetric $AdS_4$ critical
point with $SO(3)\times SO(3)\sim SO(4)_{\textrm{inv}}$ symmetry and the
corresponding dimensions of the dual operators.}\label{table3}
\end{table}

\subsection{Holographic RG flows between $N=4$ SCFTs}
\begin{figure}
         \centering
         \begin{subfigure}[b]{0.4\textwidth}
                 \includegraphics[width=\textwidth]{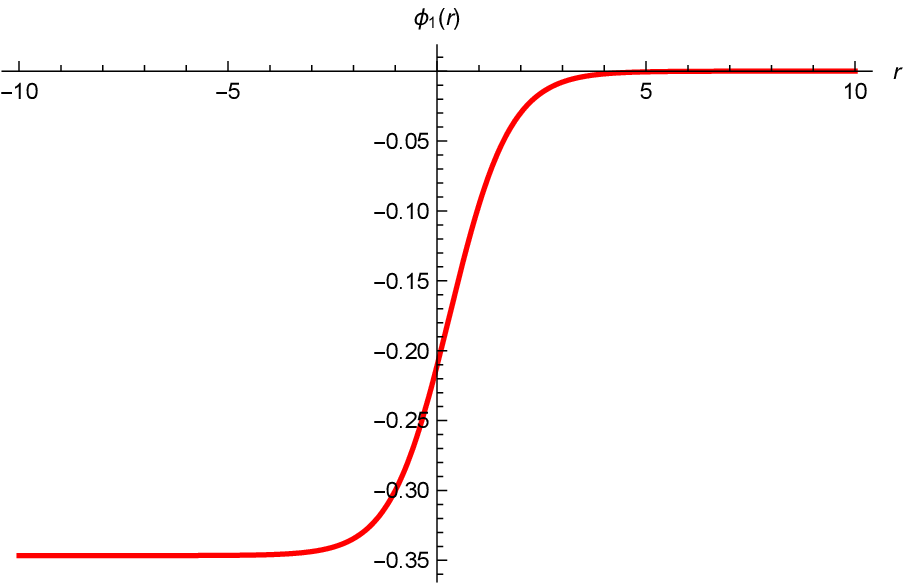}
                 \caption{Solution for $\phi_1$}
         \end{subfigure} \qquad
\begin{subfigure}[b]{0.4\textwidth}
                 \includegraphics[width=\textwidth]{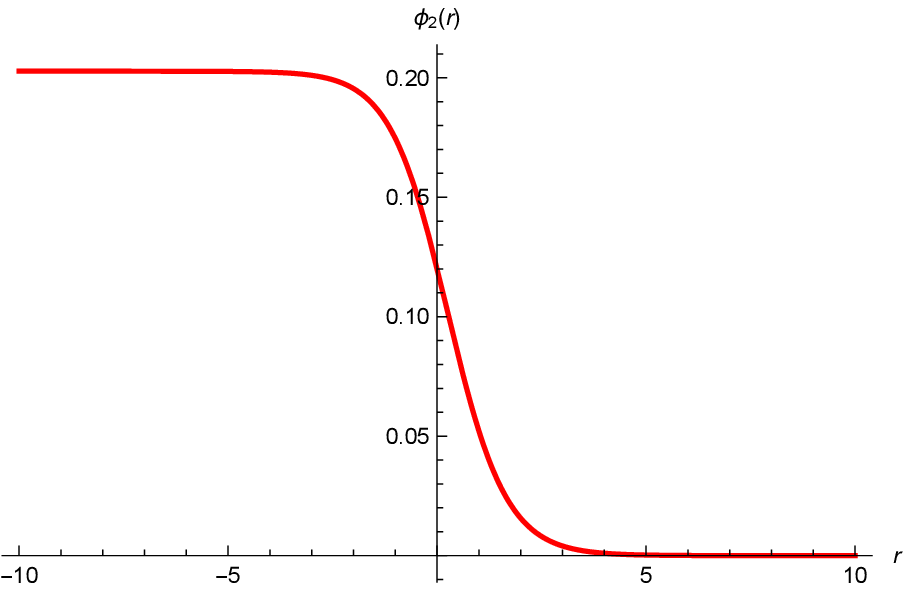}
                 \caption{Solution for $\phi_2$}
         \end{subfigure}

         ~ 
         \begin{subfigure}[b]{0.4\textwidth}
                 \includegraphics[width=\textwidth]{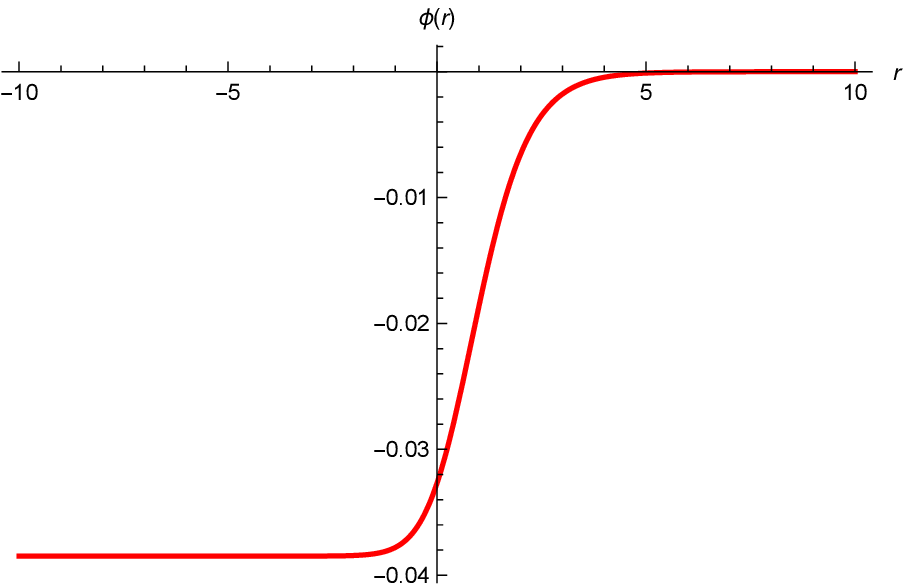}
                 \caption{Solution for $\phi$}
         \end{subfigure}\qquad 
         \begin{subfigure}[b]{0.4\textwidth}
                 \includegraphics[width=\textwidth]{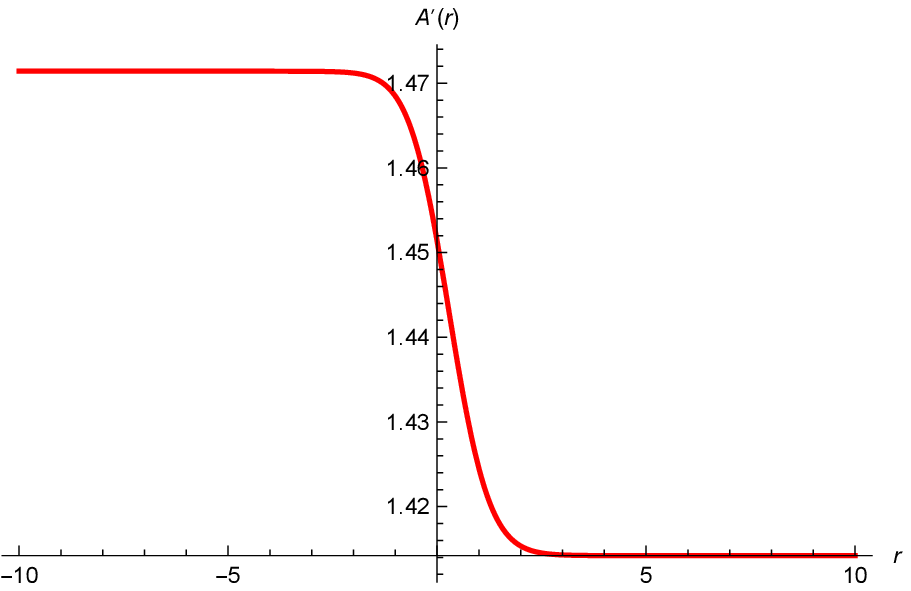}
                 \caption{Solution for $A'$}
         \end{subfigure}
         \caption{An RG flow between critical points I and IV with $g_1=1$, $\tilde{g}_1=\tilde{g}_2=2$ and $g_2=3$.}\label{fig1}
 \end{figure}

\begin{figure}
         \centering
         \begin{subfigure}[b]{0.45\textwidth}
                 \includegraphics[width=\textwidth]{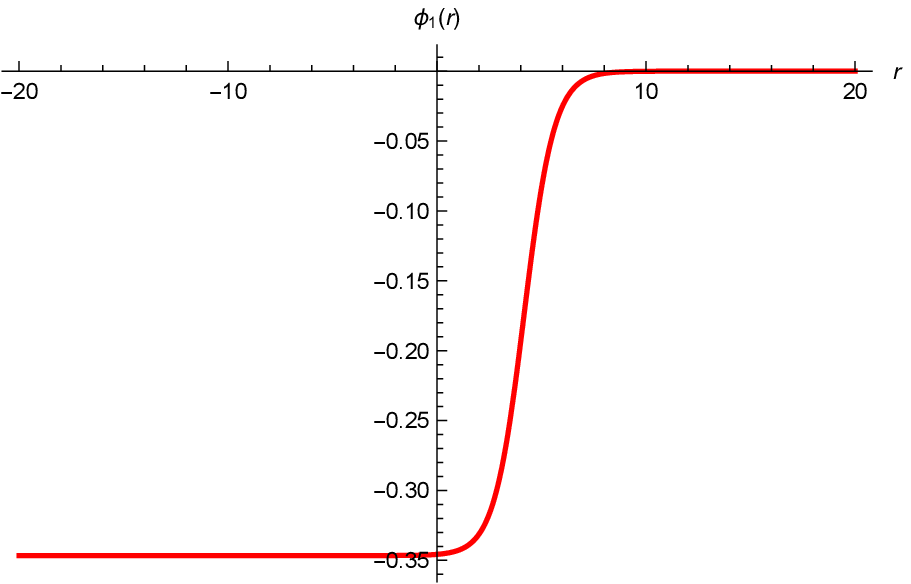}
                 \caption{Solution for $\phi_1$}
         \end{subfigure} \qquad
\begin{subfigure}[b]{0.45\textwidth}
                 \includegraphics[width=\textwidth]{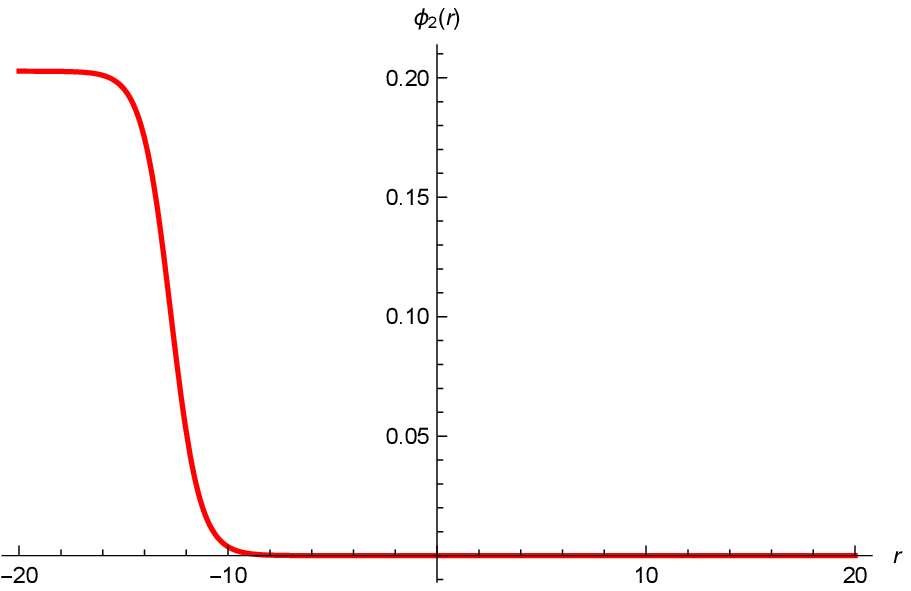}
                 \caption{Solution for $\phi_2$}
         \end{subfigure}

         ~ 
         \begin{subfigure}[b]{0.45\textwidth}
                 \includegraphics[width=\textwidth]{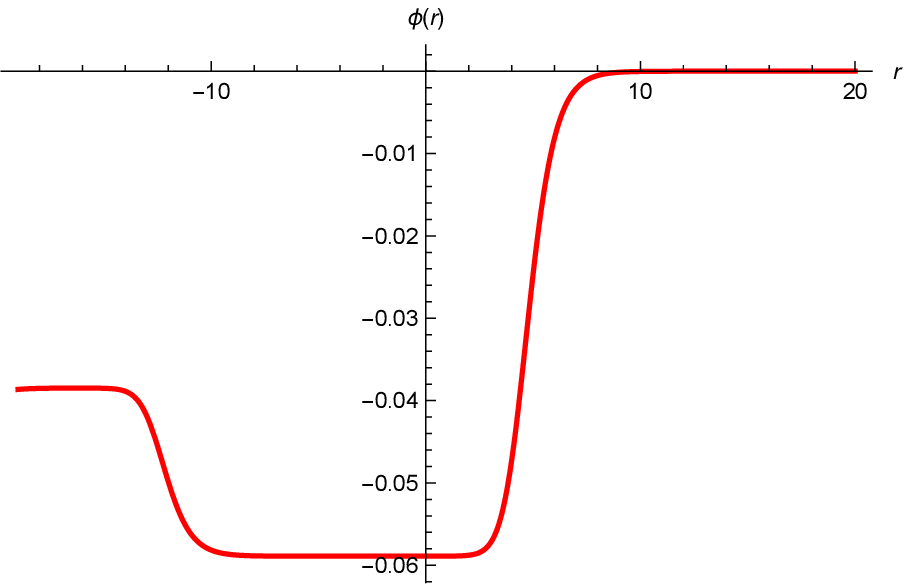}
                 \caption{Solution for $\phi$}
         \end{subfigure}\qquad 
         \begin{subfigure}[b]{0.45\textwidth}
                 \includegraphics[width=\textwidth]{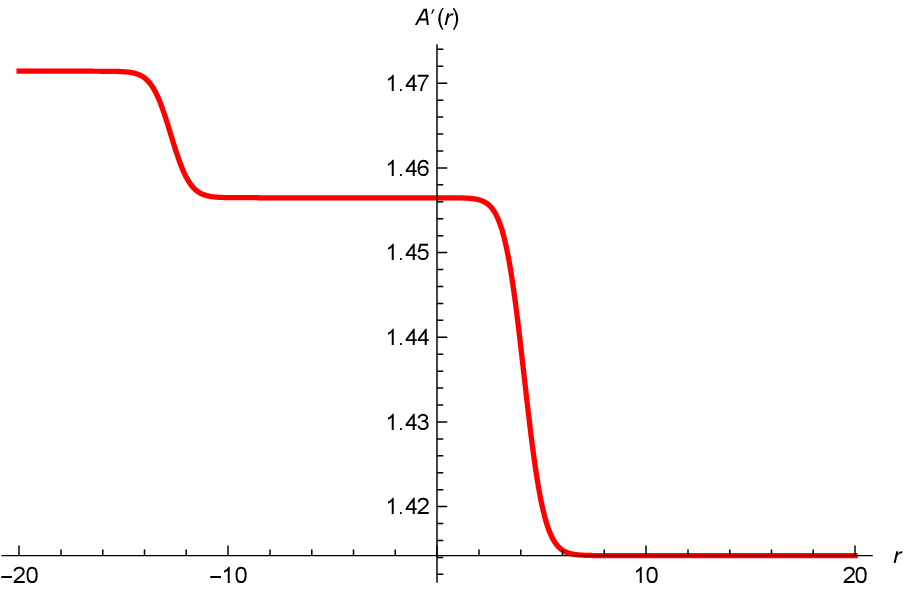}
                 \caption{Solution for $A'$}
         \end{subfigure}
         \caption{An RG flow from critical point I to critical point II and continue to critical point IV with $g_1=1$, $\tilde{g}_1=\tilde{g}_2=2$ and $g_2=3$.}\label{fig2}
 \end{figure}

We now consider holographic RG flow solutions interpolating between supersymmetric $AdS_4$ vacua previously identified. To find supersymmetric flow solutions, we begin with the metric ansatz
\begin{equation}
ds^2=e^{2A(r)}dx^2_{1,2}+dr^2
 \end{equation}
where $dx^2_{1,2}$ is the flat Minkowski metric in three dimensions.
\\
\indent For spinor conventions, we will use the Majorana representation with all $\gamma^\mu$ real and $\gamma_5$ purely imaginary. This choice implies that $\epsilon_i$ is a complex conjugate of $\epsilon^i$. All scalar fields will be functions of only the radial coordinate $r$ in order to preserve Poincare symmetry in three dimensions. The BPS conditions coming from setting $\delta\chi^i=0$ and $\delta \lambda^i_a=0$ require the following projection
\begin{equation}
\gamma_{\hat{r}}\epsilon^i=e^{i\Lambda}\epsilon_i\, .\label{gamma_r_projector}
\end{equation}
\indent It follows from the $\delta\psi_{\mu i}=0$ conditions for $\mu=0,1,2$, that
\begin{equation}
A'=\pm W,\qquad e^{i\Lambda}=\pm \frac{\mc{W}}{W}\label{A_eq_phase}
\end{equation}
where $W=|{\mc{W}}|$, and $'$ denotes the $r$-derivative. The superpotential $\mc{W}$ is defined by
\begin{equation}
\mc{W}=\frac{2}{3}\alpha
\end{equation}
where $\alpha$ is the eigenvalue of $A^{ij}_1$ corresponding to the unbroken supersymmetry. The detailed analysis leading to equation \eqref{A_eq_phase} can be found, for example, in \cite{N3_4D_gauging}. 
\\
\indent For $SO(4)_{\textrm{inv}}$ singlet scalars, the tensor $A^{ij}_1$ takes the form of a diagonal matrix
\begin{equation}
A^{ij}_1=\frac{3}{2}\mc{W}\delta^{ij}
\end{equation}
with the superpotential given by
\begin{eqnarray}
\mc{W}&=&\frac{1}{4\sqrt{2}}e^{-\frac{\phi}{2}-3\phi_1-3\phi_2}\left[3i\tilde{g}_1e^{\phi+2\phi_1+3\phi_2}+i\tilde{g}_1e^{\phi+6\phi_1+3\phi_2} \right.\nonumber \\
& &+e^{3\phi_1}(i+e^\phi\chi)[g_2(1+3e^{4\phi})-\tilde{g}_2e^{2\phi_2}(3+e^{4\phi_2})]\nonumber \\
& &\left. -ig_1e^{\phi+3\phi_2}-3ig_1e^{\phi+4\phi_1+3\phi_2}\right].\label{W_SO4_SO4}
\end{eqnarray}
\indent The variation of $\lambda^{i}_a$ leads to the following BPS equations
\begin{eqnarray}
\phi'_1&=&-\frac{i}{2\sqrt{2}}e^{i\Lambda}e^{\frac{\phi}{2}-3\phi_1}(e^{4\phi_1}-1)(e^{2\phi_1}\tilde{g}_1-g_1),\\
\phi'_2&=&-\frac{1}{2\sqrt{2}}e^{i\Lambda}e^{-\frac{\phi}{2}-3\phi_2}(e^{4\phi_2}-1)(e^{2\phi_2}\tilde{g}_2-g_2)(e^\phi\chi-i).
\end{eqnarray}
Consistency of the first condition implies that the phase $e^{i\Lambda}$ is purely imaginary, $e^{i\Lambda}=\pm i$. With this choice, the second condition requires that $\chi=0$. This is also consistent with the variation of the dilatini $\chi^i$. Furthermore, with $\chi=0$, the superpotential \eqref{W_SO4_SO4} is purely imaginary in agreement with the spinor phase $e^{i\Lambda}$ in equation \eqref{A_eq_phase}. 
\\
\indent We will choose a definite sign in order to identify the $SO(4)\times SO(4)$ critical point with the limit $r\rightarrow \infty$. The BPS equations for $\phi$, $\phi_1$ and $\phi_2$ can then be written as
\begin{equation}
\phi'=-4\frac{\pd W}{\pd \phi},\qquad \phi'_1=-\frac{2}{3}\frac{\pd W}{\pd \phi_1},\qquad \phi'_2=-\frac{2}{3}\frac{\pd W}{\pd \phi_2}
\end{equation}
together with the $A'$ equation
\begin{equation}
A'=W\, .
\end{equation}  
Explicitly, these equations read
\begin{eqnarray}
\phi'_1&=&-\frac{1}{2\sqrt{2}}e^{\frac{\phi}{2}-3\phi_1}(e^{4\phi_1}-1)(e^{2\phi_1}\tilde{g}_1-g_1),\label{phi1_eq}\\
\phi'_2&=&\frac{1}{2\sqrt{2}}e^{-\frac{\phi}{2}-3\phi_2}(e^{4\phi_2}-1)(e^{2\phi_2}\tilde{g}_2-g_2),\\
\phi'&=&-\frac{1}{2\sqrt{2}}e^{-\frac{\phi}{2}-3\phi_1-3\phi_2}\left[3\tilde{g}_1e^{\phi+2\phi_1+3\phi_2}-g_1e^{\phi+3\phi_2}-3g_1e^{\phi+4\phi_1+3\phi_2} \right.\nonumber \\
& &\left.+\tilde{g}_1e^{\phi+6\phi_1+3\phi_2}+e^{3\phi_1}[\tilde{g}_2e^{2\phi_2}(3+e^{4\phi_2})-g_2(1+3e^{4\phi_2})]
 \right],\\
A'&=&\frac{1}{4\sqrt{2}}e^{-\frac{\phi}{2}-3\phi_1-3\phi_2}\left[3\tilde{g}_1e^{\phi+2\phi_1+3\phi_2}-g_1e^{\phi+3\phi_2}-3g_1e^{\phi+4\phi_1+3\phi_2} \right.\nonumber \\
& &\left. +\tilde{g}_1e^{\phi+6\phi_1+3\phi_2}+e^{3\phi_1}[g_2(1+3e^{4\phi_2})-\tilde{g}_2e^{2\phi_2}(3+e^{4\phi_2})]\right] 
\end{eqnarray}
The scalar potential can be written in term of the real superpotential $W$ as
\begin{equation}
V=4\left(\frac{\pd W}{\pd\phi}\right)^2+\frac{2}{3}\left(\frac{\pd W}{\pd\phi_1}\right)^2+\frac{2}{3}\left(\frac{\pd W}{\pd\phi_2}\right)^2-3W^2\, .
\end{equation}
It can be verified that the above BPS equations are compatible with the second-order field equations. It should be noted that the consistency between BPS equations and field equations also requires $\chi=0$.  
\\
\indent We now consider various possible RG flows interpolating between the $N=4$ supersymmetric fixed points. Some of these flows can be obtained analytically, but the others require some sort of numerical analysis. Near the $SO(4)\times SO(4)$ critical point as $r\rightarrow \infty$, the BPS equations give 
\begin{equation}
\phi,\phi_1,\phi_2\sim e^{-\frac{r}{L_{\textrm{I}}}}
\end{equation}
in agreement with the fact that all these scalars are dual to operators of dimensions $\Delta=1,2$. $L_{\textrm{I}}$ is the $AdS_4$ radius at critical point I.
\\
\indent We begin with the flow between critical points I and II. In this case, we can consistently set $\phi_2=0$. By considering $\phi$ and $A$ as functions of $\phi_1$, we can combine the above BPS equations into
\begin{eqnarray}
\frac{d\phi}{d\phi_1}&=&-\frac{g_1(1+3e^{4\phi_1})+e^{2\phi_1}[4(g_2-\tilde{g}_2)e^{\phi_1-\phi}-\tilde{g}_1(e^{4\phi_1}+3)]}{(e^{4\phi_1}-1)(\tilde{g}_1e^{2\phi_1}-g_1)}\\
\frac{dA}{d\phi_1}&=&\frac{g_1(1+3e^{4\phi_1})-e^{2\phi_1}[4(g_2-\tilde{g}_2)e^{\phi_1-\phi}+\tilde{g}_1(3+e^{4\phi_1})]}{2(e^{4\phi_1}-1)(\tilde{g}_1e^{2\phi_1}-g_1)}\, .\label{dAdphi1}
\end{eqnarray} 
The first equation can be solved by
\begin{equation}
\phi=\ln\left[\frac{g_2-\tilde{g}_2+C_1(e^{4\phi_1}-1)}{\tilde{g}_1e^{3\phi_1}-g_1e^{\phi_1}}\right].
\end{equation}
The integration constant $C_1$ will be chosen such that the solution interpolates between $\phi=0$ at the $SO(4)\times SO(4)$ critical point and $\phi=\ln\left[\frac{\sqrt{g_1\tilde{g}_1}}{g_1+\tilde{g}_1}\right]$ at the $SO(3)_+\times SO(4)_-$ critical point. This is achieved by choosing $C_1=\frac{\tilde{g}_1^2(g_2-\tilde{g}_2)}{\tilde{g}_1^2-g_1^2}$, and the solution for $\phi$ is given by
\begin{equation}
\phi=\ln\left[\frac{(g_2-\tilde{g}_2)(g_1+\tilde{g}_1e^{2\phi_1})e^{-\phi_1}}{\tilde{g}_1^2-g_1^2}\right].
\end{equation}
With this solution, equation \eqref{dAdphi1} can be solved by
\begin{equation}
A=\frac{\phi_1}{2}-\ln(1-e^{4\phi_1})+\ln(g_1-\tilde{g}_1e^{2\phi_1})+\frac{1}{2}\ln (g_1+\tilde{g}_1e^{2\phi_1})
\end{equation}
where an irrelevant additive integration constant has been removed.
\\
\indent By changing to a new radial coordinate $\tilde{r}$ defined by $\frac{d\tilde{r}}{dr}=e^{\frac{\phi}{2}}$, equation \eqref{phi1_eq} becomes
\begin{equation}
\frac{d\phi_1}{d\tilde{r}}=-\frac{1}{2\sqrt{2}}e^{-3\phi_1}(e^{4\phi_1}-1)(\tilde{g}_1e^{2\phi_1}-g_1)
\end{equation}
whose solution is given by
\begin{eqnarray}
\frac{(g_1^2-\tilde{g}_1^2)\tilde{r}}{\sqrt{2}}&=&(g_1-\tilde{g}_1)\tan^{-1}e^{\phi_1}-(g_1+\tilde{g}_1)\tanh^{-1}e^{\phi_1}\nonumber \\
& &+2\sqrt{g_1\tilde{g}_1}\tanh^{-1}\left[\sqrt{\frac{\tilde{g}_1}{g_1}}e^{\phi_1}\right].\label{phi1_sol1_SO4}
\end{eqnarray}
Near critical point II, the operator dual to $\phi_1$ becomes irrelevant with dimension $\Delta=4$, but the operator dual to the dilaton $\phi$ is still relevant with dimensions $\Delta=1,2$. This can be seen by looking at the behavior of scalars near critical point II 
\begin{equation}
\phi\sim e^{-\frac{r}{L_{\textrm{II}}}}\qquad \textrm{and}\qquad \phi_1\sim e^{\frac{r}{L_{\textrm{II}}}}
\end{equation}
as $r\rightarrow -\infty$.
\\
\indent We can then consider a flow from critical point II to critical point IV. Along this flow, we have $\phi_1=\frac{1}{2}\ln\frac{g_1}{\tilde{g_1}}$, and a similar analysis gives the solution
\begin{eqnarray}
\phi&=&\ln\left[\frac{2\sqrt{g_1\tilde{g}_1}(g_2+\tilde{g}_2)e^{2\phi_2}}{(g_1+\tilde{g}_1)(g_2+\tilde{g}_2e^{2\phi_2})}\right],\\
A&=&\frac{\phi_2}{2}-\ln(1-e^{4\phi_2})+\ln (\tilde{g}_2e^{2\phi_2}-g_2)+\frac{1}{2}\ln(g_2+\tilde{g}_2e^{2\phi_2}).
\end{eqnarray}
Along this flow, the running of $\phi_2$ is described by
\begin{eqnarray}
\frac{(g_1-\tilde{g}_1)(g_2+\tilde{g}_2)}{\sqrt{2}}\bar{r}&=&(\tilde{g}_1-g_1)\tan^{-1}e^{\phi_2}-(g_2+\tilde{g}_2)\tanh^{-1}e^{\phi_2}\nonumber \\
& &+2\sqrt{g_2\tilde{g}_2}\tanh^{-1}\left[e^{\phi_2}\sqrt{\frac{\tilde{g}_2}{g_2}}\right]\label{phi2_sol1_SO4}
\end{eqnarray}
with $\bar{r}$ defined by $\frac{d\bar{r}}{dr}=e^{-\frac{\phi}{2}}$.
\\
\indent Similarly, the flows between critical points I and III and between critical points III and IV are given respectively by
\begin{eqnarray}
\phi_1&=&0,\\
\phi&=&\ln\left[\frac{e^{\phi_2}(g_2+\tilde{g}_2)}{g_2+e^{2\phi_2}\tilde{g}_2}\right],\\
A&=&\frac{\phi_2}{2}-\ln(1-e^{4\phi_2})+\ln (e^{2\phi_2}\tilde{g}_2-g_2)+\frac{1}{2}\ln (g_2+\tilde{g}_2e^{2\phi_2})
\end{eqnarray}
and
\begin{eqnarray}
\phi_2&=&\frac{1}{2}\ln\left[\frac{g_2}{\tilde{g}_2}\right],\\
\phi&=&\ln\left[\frac{e^{-\phi_1}(g_2+\tilde{g}_2)(g_1+\tilde{g}_1e^{2\phi_1})}{2(g_1+\tilde{g}_1)\sqrt{g_2\tilde{g}_2}}\right],\\
A&=&\frac{\phi_1}{2}-\ln(1-e^{4\phi_1})+\ln (g_1-e^{2\phi_1}\tilde{g}_1)+\frac{1}{2}\ln (g_1+\tilde{g}_1e^{2\phi_1})
\end{eqnarray}
In these cases, the $r$-dependent of $\phi_1$ and $\phi_2$ can be obtained in the same way as equations \eqref{phi1_sol1_SO4} and \eqref{phi2_sol1_SO4}.
\\
\indent For a direct flow from critical point I to critical point IV, a numerical solution is needed. This solution is given in figure \ref{fig1}. More generally, a flow from critical point I to critical point II and finally to critical point IV can also be found. This solution is given in figure \ref{fig2} and describes a cascade of RG flows with smaller flavor symmetry along the flow.    

\subsection{RG flows to $N=4$ non-conformal theory}
\begin{figure}[h]
         \centering
                 \includegraphics[width=0.75 \textwidth]{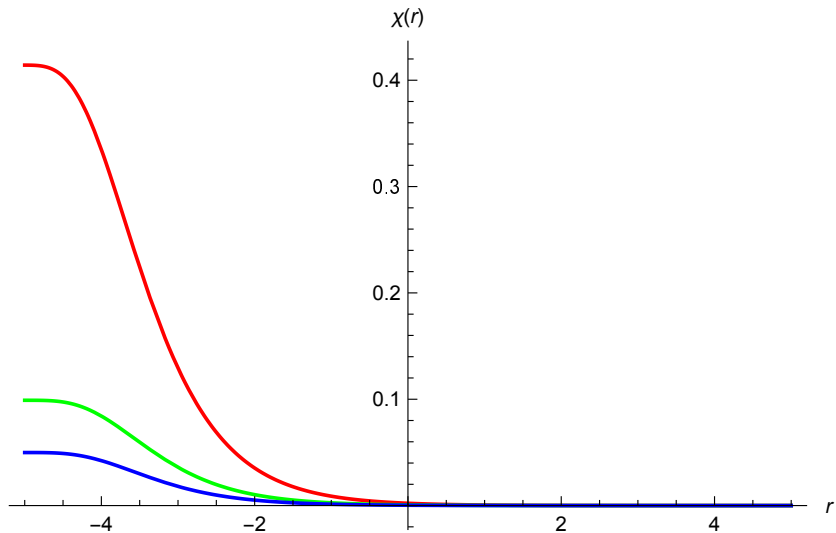}
                 \caption{Solutions for $\chi$ with $g_1=1$, $\tilde{g}_1=2$ and $\chi_0=\sqrt{1+C^2}-C$ for $C=1$ (red), C=5 (green) and C=10 (blue).}\label{fig3_SO4}
\end{figure}

A consistent truncation of the above $N=4$ $SO(4)\times SO(4)$ gauged supergravity is obtained by setting $\phi_1=\phi_2=0$. In this case, only scalars in the gravity multiplet are present. As previously mentioned, the axion $\chi$ cannot be turned on simultaneously with $\phi_1$ and $\phi_2$.   
\\
\indent For $\phi_1=\phi_2=0$, the superpotential is complex and given by
\begin{equation}
\mc{W}=\frac{1}{\sqrt{2}}e^{-\frac{\phi}{2}}\left[(g_2-\tilde{g}_2)\chi e^\phi-i(\tilde{g}_2-g_2+e^\phi(g_1-\tilde{g}_1))\right].
\end{equation}
With $\tilde{g}_2=g_1+g_2-\tilde{g}_1$, the scalar potential takes a simpler form
\begin{eqnarray}
V&=&4\left(\frac{\pd W}{\pd \phi}\right)^2+4e^{-2\phi}\left(\frac{\pd W}{\pd \chi}\right)^2-3W^2\nonumber \\
&=&-(g_1-\tilde{g}_1)^2e^{-\phi}[1+4e^\phi+e^{2\phi}(1+\chi^2)]\label{SO4_poten_gravity}
\end{eqnarray}
which has only one $AdS_4$ critical point at $\phi=\chi=0$. This is critical point I of the previous subsection. 
\\
\indent The BPS equations in this truncation are given by 
\begin{eqnarray}
\phi'&=&-4\frac{\pd W}{\pd \phi}=-\frac{\sqrt{2}(\tilde{g}_1-g_1)[e^{2\phi}(1+\chi^2)-1]}{\sqrt{(1+e^{\phi})^2+e^{2\phi}\chi^2}},\\
\chi'&=&-4e^{-2\phi}\frac{\pd W}{\pd \chi}=-\frac{2\sqrt{2}(\tilde{g}_1-g_1)e^{-\frac{\phi}{2}}\chi}{\sqrt{(1+e^\phi)^2+e^{2\phi}\chi^2}},\label{chi_eq1_SO4}\\
A'&=&W=\frac{1}{\sqrt{2}}(\tilde{g}_1-g_1)e^{-\frac{\phi}{2}}\sqrt{(1+e^\phi)^2+e^{2\phi}\chi^2}\, .
\end{eqnarray}
Near the $AdS_4$ critical point, we find
\begin{equation}
\phi\sim \chi\sim e^{-\frac{r}{L_{\textrm{I}}}}\label{phi_chi_AdS4_SO4}
\end{equation}
implying that $\phi$ and $\chi$ correspond to relevant operators of dimensions $\Delta=1,2$.
\\
\indent By considering $\phi$ and $A$ as functions of $\chi$, we can combine the BPS equations into
\begin{eqnarray}
\frac{d\phi}{d\chi}&=&\frac{e^{2\phi}(1+\chi^2)-1}{2\chi},\\
\frac{dA}{d\chi}&=&-\frac{1+2e^\phi+e^{2\phi}(1+\chi^2)}{4\chi}
\end{eqnarray}  
which can be solved by
\begin{eqnarray}
\phi&=&-\frac{1}{2}\ln(1-2C\chi-\chi^2),\label{phi_sol1_SO4}\\
A&=&-\ln\chi+\frac{1}{2}\ln[1-C\chi+\sqrt{1-2C\chi-\chi^2}]+\frac{1}{4}\ln(1-2C\chi-\chi^2)
\end{eqnarray}
in which an additive integration constant for $A$ has been neglected. It should also be noted that we must keep the constant $C\neq 0$ in order to obtain the correct behavior near the $AdS_4$ critical point as given in equation \eqref{phi_chi_AdS4_SO4}.
\\
\indent Finally, we can substitute the $\phi$ solution in \eqref{phi_sol1_SO4} in equation \eqref{chi_eq1_SO4} and in principle solve for $\chi$ as a function of $r$. However, we are not able to solve for $\chi$ analytically. We then look for numerical solutions. From equation \eqref{phi_sol1_SO4}, we see that $\phi\rightarrow 0$ as $\chi\rightarrow 0$. This limit, as usual, corresponds to the $AdS_4$ critical point. We can also see that $\phi$ is singular at $\chi_0$ for which $1-2C\chi_0 -\chi_0^2=0$ or
\begin{equation}
\chi_0=-C\pm \sqrt{1+C^2}\, .
\end{equation}
This implies that $\phi$ flows from the value $\phi=0$ at the critical point to a singular value $\phi\rightarrow \infty$ while $\chi$ flows between the values $\chi=0$ and $\chi=\chi_0$. Examples of solutions for $\chi$ is shown in figure \ref{fig3_SO4}. 
\\
\indent Near the singularity $\phi\rightarrow \infty$ and $\chi\rightarrow \chi_0$, we find that 
\begin{equation}
\chi-\chi_0\sim r^4,\qquad \phi\sim -\ln r^2,\qquad A\sim \ln r\, .
\end{equation}
This gives the metric
\begin{equation}
ds^2=r^2dx^2_{1,2}+dr^2\, .
\end{equation}
From the scalar potential \eqref{SO4_poten_gravity}, we find $V\rightarrow -\infty$ for any value of $\chi_0$. Therefore, the singularity is physical according to the criterion of \cite{Gubser_singularity}. We then conclude that the solution describes an RG flow from the $N=4$ SCFT in the UV to a non-conformal field theory in the IR corresponding to the above singularity. The deformations break conformal symmetry but preserve the $SO(4)$ flavor symmetry and $N=4$ Poincare supersymmetry in three dimensions.

\section{$N=4$ $SO(3,1)\times SO(3,1)$ gauged supergravity}\label{SO3_1_SO3_1}
In this section, we consider non-compact gauge group $SO(3,1)\times SO(3,1)$ with the embedding tensor
\begin{eqnarray}
& &f_{+123}=f_{+189}=f_{+729}=-f_{+783}=\frac{1}{\sqrt{2}}(g_1-\tilde{g}_1),\nonumber \\ 
& &f_{+789}=f_{+183}=f_{+723}=-f_{+129}=\frac{1}{\sqrt{2}}(g_1+\tilde{g}_1),\nonumber \\
& &f_{-456}=f_{-4,11,12}=f_{-10,5,12}=-f_{-10,11,6}=\frac{1}{\sqrt{2}}(g_2-\tilde{g}_2),\nonumber \\
& &f_{-10,11,12}=f_{-4,11,6}=f_{-10,5,6}=-f_{-45,12}=\frac{1}{\sqrt{2}}(g_2+\tilde{g}_2).
\end{eqnarray}
We now repeat the analysis performed in the previous section.

\subsection{Supersymmetric $AdS_4$ vacuum}
We will parametrize the $SO(6,6)/SO(6)\times SO(6)$ coset by using scalars that are $SO(3)\times SO(3)\subset SO(3,1)\times SO(3,1)$ invariant. From the embedding of $SO(3,1)$ in $SO(3,3)$, there are two $SO(3)\times SO(3)$ singlets corresponding to the non-compact generators 
\begin{equation}
\tilde{Y}_1=Y_{11}+Y_{22}-Y_{33},\qquad \tilde{Y}_2=Y_{44}+Y_{55}-Y_{66}\, .\label{SO3_1_non_compact}
\end{equation}
The coset representative can be parametrized as
\begin{equation}
L=e^{\phi_1\tilde{Y}_1}e^{\phi_2\tilde{Y}_2}\, .
\end{equation}
\indent In this case, the scalar potential is given by
\begin{eqnarray}
V&=&\frac{1}{8}e^{-\phi-6\phi_1-6\phi_2}\left[2g_2e^{\phi+3\phi_1+9\phi_2}(e^{6\phi_1}g_1-3g_1e^{2\phi_1}-\tilde{g}_1+3\tilde{g}_1e^{4\phi_1}) \right.\nonumber \\
& &-6g_2e^{\phi+3\phi_1+5\phi_2}(g_1e^{6\phi_1}-3g_1e^{2\phi_1}-\tilde{g}_1+3\tilde{g}_1e^{4\phi_1})\nonumber \\
& &+6\tilde{g}_2e^{\phi+3\phi_1 +7\phi_2}(3\tilde{g}_1e^{4\phi_1}-3g_1e^{2\phi_1}+g_1e^{6\phi_1}-\tilde{g}_1)\nonumber \\
& &-2\tilde{g}_2e^{\phi+3\phi_1+3\phi_2}(g_1e^{6\phi_1}-3g_1e^{2\phi_1}-\tilde{g}_1+3\tilde{g}_1e^{4\phi_1})\nonumber \\
& &+3e^{6\phi_1+4\phi_2}[e^{4\phi_2}(2g_2^2-\tilde{g}_2^2)(1+\chi^2e^{2\phi})-3(g_2^2-2\tilde{g}_2^2)(1+\chi^2e^{2\phi})]\nonumber \\
& &+g_2^2e^{6\phi_1+6\phi_2}(1+\chi^2e^{2\phi})+\tilde{g}_2^2e^{6\phi_1}(1+\chi^2e^{2\phi})\nonumber \\
& &+e^{6\phi_2}\left[3(2g_1^2-\tilde{g}_1^2)e^{2\phi+8\phi_1}+16e^{6\phi_1}[g_2\tilde{g}_2+e^{2\phi}(g_1\tilde{g}_1+g_2\tilde{g}_2\chi^2)] \right.\nonumber \\
& &\left. \left.+g_1^2e^{2\phi+12\phi_1}\tilde{g}_1^2e^{2\phi}-3(g_1^2-2\tilde{g}_1^2)e^{\phi+4\phi_1} \right]\right].
\end{eqnarray}
\indent This potential admits only one supersymmetric $AdS_4$ critical point at
\begin{eqnarray}
\phi&=&\frac{1}{2}\ln\left[\frac{g_1\tilde{g}_1(g_2^2+\tilde{g}_2^2)^2(g_1^2+\tilde{g}_1^2)^2}{g_2\tilde{g}_2}\right], \qquad \chi=0,\nonumber \\
\phi_1&=&\frac{1}{2}\ln \left[-\frac{\tilde{g}_1}{g_1}\right],\qquad \phi_2=\frac{1}{2}\ln \left[-\frac{\tilde{g}_2}{g_2}\right].
\end{eqnarray}
This critical point preserves $N=4$ supersymmetry and $SO(3)\times SO(3)$ symmetry. The latter is the maximal compact subgroup of $SO(3,1)\times SO(3,1)$ gauge group. Without loss of generality, we can shift the scalars such that the critical point occurs at $\phi=\phi_1=\phi_2=0$. This can be achieved by setting
\begin{equation}
\tilde{g}_1=-g_1,\qquad \tilde{g}_2=-g_2,\qquad g_2=-g_1\, .
\end{equation} 
With these values, the cosmological constant and $AdS_4$ radius are given by
\begin{equation}
V_0=-6g_1^2\qquad \textrm{and}\qquad L^2=\frac{1}{2g_1^2}\, .
\end{equation}
It should be noted that the choice $\tilde{g}_1=-g_1$, $\tilde{g}_2=-g_2$ and $g_2=g_1$ makes the critical point at $\phi=\phi_1=\phi_2=\chi=0$ a $dS_4$ with $V_0=2g_1^2$.
\\
\indent At the $N=4$ $AdS_4$ critical point, the gauge group $SO(3,1)\times SO(3,1)$ is broken down to its maximal compact subgroup $SO(3)\times SO(3)$. All scalar masses at this critical point are given in table \ref{table4}. The two singlet representations $(\mathbf{1},\mathbf{1})$ corresponding to $\phi_1$ and $\phi_2$ are dual to irrelevant operators of dimensions $\Delta=4$, and six massless scalars in representation $(\mathbf{1},\mathbf{3})+(\mathbf{3},\mathbf{1})$ are Goldstone bosons.

\begin{table}[h]
\centering
\begin{tabular}{|c|c|c|}
  \hline
  Scalar field representations & $m^2L^2\phantom{\frac{1}{2}}$ & $\Delta$  \\ \hline
  $(\mathbf{1},\mathbf{1})$ & $-2_{\times 2}$ &  $1,2$  \\
  $(\mathbf{1},\mathbf{1})$ & $4_{\times 2}$ &  $4$  \\
   $(\mathbf{1},\mathbf{5})+(\mathbf{5},\mathbf{1})$ & $-2_{\times 10}$ &  $1,2$  \\
  $(\mathbf{1},\mathbf{3})+(\mathbf{3},\mathbf{1})$ & $0_{\times 6}$ &  $3$  \\
  $(\mathbf{3},\mathbf{3})$ & $0_{\times 18}$ &  $3$  \\
  \hline
\end{tabular}
\caption{Scalar masses at the $N=4$ supersymmetric $AdS_4$ critical
point with $SO(3)\times SO(3)$ symmetry and the
corresponding dimensions of the dual operators for $SO(3,1)\times SO(3,1)$ gauge group.}\label{table4}
\end{table}

\subsection{RG flows without vector multiplet scalars}  
\begin{figure}[h]
         \centering
                 \includegraphics[width=0.6 \textwidth]{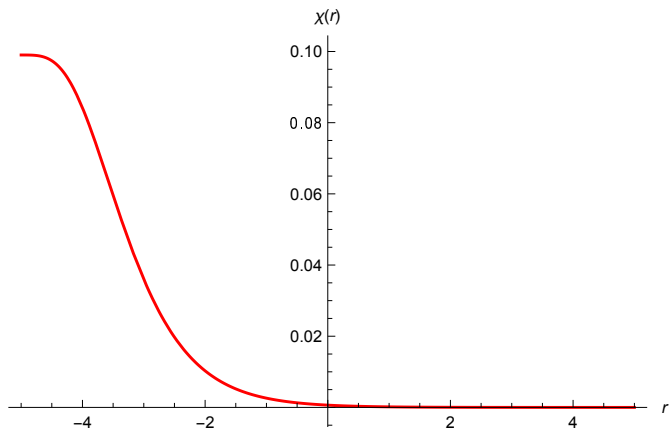}
                 \caption{Solution for $\chi$ with $\chi_0=\sqrt{1+C^2}-C$ for $C=5$ and $g_1=1$ in $SO(3,1)\times SO(3,1)$ gauging.}\label{fig1_SO31}
\end{figure}

Since there is only one supersymmetric $AdS_4$ critical point, there is no supersymmetric RG flow between the dual SCFTs. In this case, we instead consider RG flows from the SCFT dual to the $N=4$ $AdS_4$ vacuum with $SO(3)\times SO(3)$ symmetry. We begin with a simple truncation to scalar fields in the supergravity multiplet obtained by setting $\phi_1=\phi_2=0$. Within this truncation, the superpotential is given by
\begin{equation}
\mc{W}=\frac{3i}{2\sqrt{2}}g_1e^{-\frac{\phi}{2}}[1+e^\phi (1-i\chi)]
\end{equation}        
in term of which the scalar potential can be written as
\begin{eqnarray}
V&=&\frac{16}{9}\left(\frac{\pd W}{\pd \phi}\right)^2+\frac{16}{9}e^{-2\phi}\left(\frac{\pd W}{\pd \chi}\right)^2-\frac{4}{3}W^2\nonumber \\
&=&-g_1^2e^{-\phi}[1+4e^{\phi}+e^{2\phi}(1+\chi^2)].
\end{eqnarray}   
\indent The flow equations obtained from $\delta \chi^i=0$ conditions are given by
\begin{eqnarray}
\phi'&=&-\frac{8}{3}\frac{\pd W}{\pd \phi}=-\frac{\sqrt{2}g_1e^{-\frac{\phi}{2}}[e^{2\phi}(1+\chi^2)-1]}{\sqrt{(1+e^\phi)^2+e^{2\phi}\chi^2}},\label{phi_eq1_SO31}\\
\chi'&=&-\frac{8}{3}e^{-2\phi}\frac{\pd W}{\pd \chi}=-\frac{2\sqrt{2}g_1e^{-\frac{\phi}{2}}\chi}{\sqrt{(1+e^\phi)^2+e^{2\phi}\chi^2}}\, .\label{chi_eq1_SO31}
\end{eqnarray}
The BPS conditions from $\delta \lambda^i_a=0$ are, of course, identically satisfied by setting $\phi_1=\phi_2=0$. 
\\     
\indent The flow equation for the metric function is simply given by
\begin{equation}
A'=W=\frac{3}{2\sqrt{2}}g_1e^{-\frac{\phi}{2}}\sqrt{(1+e^\phi)^2+e^{2\phi}\chi^2}\, .\label{A_eq1_SO31}
\end{equation}
\indent Near the $AdS_4$ critical point, we find
\begin{equation} 
\phi\sim \chi\sim e^{-\sqrt{2}g_1r}\sim e^{-\frac{r}{L}}\label{Scalar_near_AdS4_SO13}
\end{equation}
as expected for the dual operators of dimensions $\Delta=1,2$. 
\\
\indent Apart from some numerical factors involving gauge coupling constants, the structure of the resulting BPS equations are very similar to the $SO(4)\times SO(4)$ case. We therefore only give the solution without going into any details here  
\begin{eqnarray}
\phi&=&-\frac{1}{2}\ln (1-\chi^2-2C\chi),\label{phi_sol1_SO13}\\
A&=&-\frac{3}{2}\ln \chi+\frac{3}{8}\ln(1-2C\chi-\chi^2)+\frac{3}{4}\ln (1-C\chi+\sqrt{1-2C\chi-\chi^2}).\,\,\,
\end{eqnarray}
\indent As in the $SO(4)\times SO(4)$ case, we are able to solve for $\chi$ only numerically. An example of solutions for $\chi$ is shown in figure \ref{fig1_SO31}. From the figure, it can be readily seen that, along the flow, $\chi$ interpolates between $\chi=0$ and $\chi_0=-C\pm \sqrt{1+C^2}$. The $\phi$ solution, on the other hand, interpolates between $\phi=0$ and $\phi\rightarrow \infty$ as can be seen from the solution \eqref{phi_sol1_SO13}. 
The singularity $\phi\rightarrow \infty$ also gives rise to $V\rightarrow -\infty$ for any value of $\chi_0$. Therefore, the singularity is physical, and the solution describes an RG flow from the $N=4$ SCFT in the UV with $SO(3)\times SO(3)$ symmetry to a non-conformal field theory in the IR corresponding to this singularity.

\subsection{RG flows with vector multiplet scalars} 
We now consider solutions with non-vanishing vector multiplet scalars. In this case, we need to set $\chi=0$ in order to make the solutions of the BPS equations solve the second-order field equations as in the case of $SO(4)\times SO(4)$ gauging. The corresponding BPS equations are given by 
\begin{eqnarray}
\phi'_1&=&\sqrt{2}g_1e^{\frac{\phi}{2}}\cosh(2\phi_1)\sinh \phi_1 ,\\
\phi'_2&=&\sqrt{2}g_1e^{-\frac{\phi}{2}}\cosh(2\phi_2)\sinh \phi_2,\\
\phi'&=&\sqrt{2}g_1e^{-\frac{\phi}{2}}\left[e^\phi \cosh \phi_1(\cosh(2\phi_1)-2)-\cosh \phi_2(\cosh(2\phi_2)-2) \right],\\
A'&=&\frac{1}{\sqrt{2}}g_1e^{-\frac{\phi}{2}}\left[e^\phi\cosh\phi_1 (\cosh(2\phi_1)-2)+\cosh\phi_2(\cosh(2\phi_2)-2)\right].
\end{eqnarray}
With suitable boundary conditions, these equations can be solved numerically as in the previous cases. We will, however, look at particular truncations for which analytic solutions can be found. These solutions should be more interesting and more useful than the numerical ones in many aspects.
\\
\indent The first truncation is obtained by setting $\phi_2=\phi_1$ and $\phi=0$. It can be easily verified that this is a consistent truncation. The relevant BPS equations read
\begin{eqnarray}
\phi_1'&=&\sqrt{2}g_1\cosh(2\phi_1)\sinh\phi_1,\\
A'&=&\sqrt{2}g_1\cosh\phi_1 [\cosh(2\phi_1)-2]
\end{eqnarray} 
which have a solution
\begin{eqnarray}
2g_1r&=&\ln\left[\frac{1-\sqrt{2}\cosh\phi_1}{1+\sqrt{2}\cosh\phi_1}\right]-2\sqrt{2}\tanh^{-1}e^{\phi_1},\\
A&=&\ln(1+e^{4\phi_1})-\ln(1-e^{2\phi_1})-\phi_1\, .
\end{eqnarray}
The solution for $\phi_1$ is clearly seen to be singular at a finite value of $r$. 
\\
\indent Recall that $\phi_1$ and $\phi_2$ are dual to irrelevant operators, we expect that in this case, the $N=4$ SCFT should appear in the IR. Near the singularity, we find
\begin{equation} 
\phi_1\sim \pm\frac{1}{3}\ln\left[C-\frac{3g_1r}{2\sqrt{2}}\right]\qquad \textrm{and}\qquad A\sim-\frac{1}{3}\ln\left[C-\frac{3g_1r}{2\sqrt{2}}\right]
\end{equation}
for a constant $C$. It can be verified that, in this limit, the scalar potential blows up as $V\rightarrow \infty$. Therefore, the singularity is unphysical. 
\\
\indent Another truncation is obtained by setting $\phi_2=0$ which gives rise to the BPS equations 
\begin{eqnarray}
\phi_1'&=&\sqrt{2}g_1e^{\frac{\phi}{2}}\cosh(2\phi_1)\sinh\phi_1,\\
\phi'&=&\sqrt{2}g_1e^{-\frac{\phi}{2}}(1+e^{\phi}\cosh\phi_1)[\cosh(2\phi_1)-2],\\
A'&=&\frac{1}{\sqrt{2}}g_1e^{-\frac{\phi}{2}}[e^\phi\cosh\phi_1(\cosh(2\phi_1)-2)-1].
\end{eqnarray}
An analytic solution to these equations is given by 
\begin{eqnarray}
\phi&=&\ln\left[\cosh\phi_1-\frac{1}{2}C\cosh(2\phi_1)\textrm{csch}\phi_1\right],\\
\sqrt{2}g_1\tilde{r}&=&\ln[C-\tanh(2\phi_1)],\\
A&=&\ln[\cosh(2\phi_1)]-\frac{1}{2}\ln(\sinh\phi_1)\nonumber \\
& & -\frac{1}{2}\ln[C\cosh(2\phi_1)-\sinh(2\phi_1)]
\end{eqnarray} 
where the coordinate $\tilde{r}$ is defined via $\frac{d\tilde{r}}{dr}=e^{-\frac{\phi}{2}}$. It should be noted that to give the correct behavior for $\phi$ and $\phi_1$ near the $AdS_4$ critical point, we need $C\neq 0$.
\\
\indent The solution is singular at a finite value of $\tilde{r}$. Near this singularity, we find
\begin{equation}
\phi_1\sim \pm \frac{1}{4}\ln \left |\sqrt{2}g_1\tilde{r}-\tilde{C}\right |
\end{equation}
where $\tilde{C}$ is a constant. The behavior of $\phi$ and $A$ depends on the value of $C$. 
\\
\indent We begin with the case $\phi_1\rightarrow \infty$. For $C=2$, we find from the explicit solution that
\begin{eqnarray}
\phi &\sim& -\phi_1\sim \frac{1}{4}\ln \left |\sqrt{2}g_1\tilde{r}-\tilde{C}\right |,\nonumber \\
 A&\sim& \phi_1\sim -\frac{1}{4}\ln \left |\sqrt{2}g_1\tilde{r}-\tilde{C}\right |\, . 
\end{eqnarray} 
For $C\neq 2$, we find 
\begin{eqnarray}
\phi &\sim& \phi_1\sim -\frac{1}{4}\ln \left |\sqrt{2}g_1\tilde{r}-\tilde{C}\right |,\nonumber \\
 A&\sim& \phi_1\sim -\frac{1}{4}\ln \left |\sqrt{2}g_1\tilde{r}-\tilde{C}\right |\, . 
\end{eqnarray} 
Both of these singularities lead to $V\rightarrow \infty$ and hence are unphysical. 
\\
\indent We now move to another possibility with $\phi_1\rightarrow -\infty$. In this case, we find
\begin{eqnarray}
\phi &\sim& \phi_1\sim \frac{1}{4}\ln \left |\sqrt{2}g_1\tilde{r}-\tilde{C}\right |,\nonumber \\
 A&\sim& -\phi_1\sim -\frac{1}{4}\ln \left |\sqrt{2}g_1\tilde{r}-\tilde{C}\right |
\end{eqnarray}
for $C=-2$ and 
\begin{eqnarray}
\phi &\sim& -\phi_1\sim -\frac{1}{4}\ln \left |\sqrt{2}g_1\tilde{r}-\tilde{C}\right |,\nonumber \\
 A&\sim& -\phi_1\sim -\frac{1}{4}\ln \left |\sqrt{2}g_1\tilde{r}-\tilde{C}\right |
\end{eqnarray}
for $C\neq -2$. These behaviors also give $V\rightarrow \infty$. Therefore, we conclude that the solutions in this particular truncation do not holographically describe RG flows from $N=4$ SCFT.  
\\
\indent A similar analysis shows that the truncation with $\phi_1=0$ also leads to unphysical singularites. It would be interesting to uplift these solutions to ten or eleven dimensions and determine whether these singularities are resolved.                                                                                 
                                                                                                                                                                                                 
\section{$N=4$ $SO(4)\times SO(3,1)$ gauged supergravity}\label{SO4_SO3_1}
In this section, we consider a gauge group with one compact and one non-compact factors of the form $SO(4)\times SO(3,1)$. All the procedures are essentially the same, so we will not present much detail here. The $SO(4)$ and $SO(3,1)$ are electrically and magnetically embedded in $SO(3,3)\times SO(3,3)$, respectively. The corresponding embedding tensor is given by
\begin{eqnarray}
& &f_{+123}=\sqrt{2}(g_1-\tilde{g}_1),\qquad f_{+789}=\sqrt{2}(g_1+\tilde{g}_1),\nonumber \\
& &f_{-456}=f_{-4,11,12}=f_{-10,5,12}=-f_{-10,11,6}=\frac{1}{\sqrt{2}}(g_2-\tilde{g}_2),\nonumber \\
& &f_{-10,11,12}=f_{-4,11,6}=f_{-10,5,6}=-f_{-45,12}=\frac{1}{\sqrt{2}}(g_2+\tilde{g}_2).
\end{eqnarray}

\subsection{Supersymmetric $AdS_4$ vacua}
We consider scalar fields invariant under $SO(4)_{\textrm{inv}}\subset SO(4)\times SO(3)\subset SO(4)\times SO(3,1)$. The corresponding coset representative for the $SO(6,6)/SO(6)\times SO(6)$ coset is now given by 
\begin{equation}
L=e^{\phi_1\hat{Y}}e^{\phi_2\tilde{Y}_2}
\end{equation}
where $\hat{Y}_1$ and $\tilde{Y}_2$ are defined in \eqref{SO4_inv_L} and \eqref{SO3_1_non_compact}, respectively. 
\\
\indent The scalar potential turns out to be
\begin{eqnarray}
V&=&\frac{1}{8}e^{-\phi-6\phi_1-6\phi_2}\left[(g_1+g_2)^2e^{2\phi+12\phi_1+6\phi_2}-3(3g_1^2+2g_1g_2+g_2^2)e^{2\phi+4\phi_1+6\phi_2}\right.\nonumber \\
& &+e^{6\phi_1}\left[g_2^2(1+e^{2\phi}\chi^2)(1+e^{4\phi_2})^3+16e^{6\phi_2}[e^{2\phi}(g_1^2+g_1g_2-g_2^2\chi^2)-g_2^2] \right]\nonumber \\
& &+8g_2e^{\phi+3\phi_1+6\phi_2}[g_1(e^{2\phi_1}-1)^3+g_2e^{2\phi_1}(3+e^{4\phi_1})]\cosh\phi_2\times \nonumber \\
& & \left. [\cosh(2\phi_2)-2]-3(3g_1^2+4g_1g_2+2g_2^2)e^{2\phi+8\phi_1+6\phi_2}+g_1^2e^{2\phi+6\phi_2}\right]
\end{eqnarray}
where we have imposed the following relations 
\begin{equation}
\tilde{g}_1=g_1+g_2\qquad \textrm{and}\qquad \tilde{g}_2=-g_2
\end{equation}
in order to have an $N=4$ supersymmetric $AdS_4$ critical point with $SO(4)\times SO(3)$ symmetry at $\phi_1=\phi_2=\phi=\chi=0$. 
\\
\indent There are two supersymmetric $AdS_4$ vacua with $N=4$ supersymmetry:
\begin{itemize}
\item The first critical point is a trivial one with $SO(4)\times SO(3)$ symmetry at
\begin{equation}
\phi=\chi=\phi_1=\phi_2=0,\qquad V_0=-6g_2^2\, .
\end{equation}
\item A non-trivial supersymmetric critical point is given by
\begin{eqnarray}
\phi_2&=&\chi=0,\qquad \phi_1=\frac{1}{2}\ln\left[\frac{g_1}{g_1+g_2}\right],\nonumber \\
\phi&=&\frac{1}{2}\ln\left[\frac{4g_1(g_1+g_2)}{(2g_1+g_2)^2}\right],\qquad V_0=-\frac{3g_2^2(2g_1+g_2)}{\sqrt{g_1(g_1+g_2)}}\, . 
\end{eqnarray}
This critical point is invariant under a smaller symmetry $SO(3)\times SO(3)$.
\end{itemize}
Scalar masses at these two critical points are given in tables \ref{table5} and \ref{table6}. It can be seen that the mass spectra are very similar to critical points III and IV in the case of $SO(4)\times SO(4)$ gauge group.

\begin{table}[h]
\centering
\begin{tabular}{|c|c|c|}
  \hline
  Scalar field representations & $m^2L^2\phantom{\frac{1}{2}}$ & $\Delta$  \\ \hline
  $(\mathbf{1},\mathbf{1},\mathbf{1})$ & $-2_{\times 2}$ &  $1,2$  \\
  $(\mathbf{1},\mathbf{1},\mathbf{1})$ & $4$ &  $4$  \\
   $(\mathbf{1},\mathbf{1},\mathbf{3})$ & $0_{\times 3}$ &  $3$  \\
    $(\mathbf{3},\mathbf{3},\mathbf{1})$ & $0_{\times 9}$ &  $3$  \\
      $(\mathbf{1},\mathbf{1},\mathbf{5})$ & $-2_{\times 5}$ &  $1,2$  \\
   $(\mathbf{1},\mathbf{3},\mathbf{3})+(\mathbf{3},\mathbf{1},\mathbf{3})$ & $-2_{\times 18}$ &  $1,2$  \\
  \hline
\end{tabular}
\caption{Scalar masses at the $N=4$ supersymmetric $AdS_4$ critical
point with $SO(4)\times SO(3)$ symmetry and the
corresponding dimensions of the dual operators for $SO(4)\times SO(3,1)$ gauge group.}\label{table5}
\end{table}

\begin{table}[h]
\centering
\begin{tabular}{|c|c|c|}
  \hline
  Scalar field representations & $m^2L^2\phantom{\frac{1}{2}}$ & $\Delta$  \\ \hline
  $(\mathbf{1},\mathbf{1})$ & $-2_{\times 2}$ &  $1,2$  \\
  $(\mathbf{1},\mathbf{1})$ & $4_{\times 2}$ &  $4$  \\
   $(\mathbf{1},\mathbf{5})+(\mathbf{5},\mathbf{1})$ & $-2_{\times 10}$ &  $1,2$  \\
  $(\mathbf{1},\mathbf{3})+(\mathbf{3},\mathbf{1})$ & $0_{\times 6}$ &  $3$  \\
  $(\mathbf{3},\mathbf{3})$ & $0_{\times 18}$ &  $3$  \\
  \hline
\end{tabular}
\caption{Scalar masses at the $N=4$ supersymmetric $AdS_4$ critical
point with $SO(3)\times SO(3)$ symmetry and the
corresponding dimensions of the dual operators for $SO(4)\times SO(3,1)$ gauge group.}\label{table6}
\end{table}

\subsection{Holographic RG flow}
In this section, we will give a supersymmetric RG flow solution interpolating between the two $AdS_4$ vacua identified above. As in the previous cases, turning on vector multiplet scalars requires the vanishing of the axion $\chi$. Since we are only interested in the solution interpolating between two $AdS_4$ vacua, we will accordingly set $\chi=0$ from now on. 
\\
\indent With $\chi=0$, the superpotential is given by
\begin{eqnarray}
\mc{W}&=&\frac{i}{4\sqrt{2}}e^{-\frac{\phi}{2}-3\phi_1-3\phi_2}\left[g_1e^{\phi+3\phi_2}+3g_1e^{\phi+4\phi_1+3\phi_2}-3(g_1+g_2)e^{\phi+2\phi_1+3\phi_2} \right.\nonumber \\
& &\left. +g_2e^{3\phi_1}(1+e^{2\phi_2})(1-4e^{2\phi_2}+e^{4\phi_2})-(g_1+g_2)e^{\phi+6\phi_1+3\phi_2}\right]
\end{eqnarray}
in term of which the scalar potential can be written as
\begin{equation}
V=4\left(\frac{\pd W}{\pd \phi}\right)^2+\frac{2}{3}\left(\frac{\pd W}{\pd \phi_1}\right)^2+\frac{2}{3}\left(\frac{\pd W}{\pd \phi_2}\right)^2-3W^2\, .
\end{equation}
\indent The BPS equations read
\begin{eqnarray}
\phi'_1&=&-\frac{2}{3}\frac{\pd W}{\pd \phi_1}=-\frac{1}{2\sqrt{2}}e^{\frac{\phi}{2}-3\phi_1}(e^{4\phi_1}-1)(e^{2\phi_1}(g_1+g_2)-g_1),\\
\phi'_2&=&-\frac{2}{3}\frac{\pd W}{\pd \phi_2}=\frac{1}{2\sqrt{2}}g_2e^{-\frac{\phi}{2}-3\phi_1}(e^{2\phi_2}-1)(e^{4\phi_1}+1),\\
\phi'&=&-4\frac{\pd W}{\pd \phi}=-\frac{1}{2\sqrt{2}}e^{-\frac{\phi}{2}-3\phi_1}\left[4g_2e^{3\phi_1}\cosh\phi_2[\cosh(2\phi_2)-2]\right.\nonumber \\
& &\left.+e^{\phi}\left[[(e^{2\phi_1}-1)^3g_1+e^{2\phi_1}(3+e^{4\phi_1})g_2]\right] \right],\\
A'&=&\frac{1}{4}\sqrt{2}e^{-\frac{\phi}{2}-3\phi_1}\left[e^{\phi}[(e^{2\phi_1}-1)^3g_1+e^{2\phi_1}(3+e^{4\phi_1})g_2]\right.\nonumber \\
& &\left.-4g_2e^{3\phi_1}\cosh\phi_2[\cosh(2\phi_2)-2] \right].
\end{eqnarray} 
\indent Since $\phi_2=0$ at both critical points, we can consistently truncate $\phi_2$ out. Note also that $\phi_2$ is dual to an irrelevant operator of dimension $\Delta=4$ as can be seen from the linearized BPS equations which give
\begin{equation}
\phi\sim \phi_1\sim e^{-\frac{r}{L}},\qquad \phi_2\sim e^{\frac{r}{L}}\, .
\end{equation}
With $\phi_2=0$, we find an RG flow solution driven by $\phi$ and $\phi_1$ as follow
\begin{eqnarray}
g_2(2g_1+g_2)\tilde{r}&=&\sqrt{2}g_2\tan^{-1}e^{\phi_1}+\sqrt{2}(2g_1+g_2)\tanh^{-1}e^{\phi_1}\nonumber \\
& &-2\sqrt{2g_1(g_1+g_2)}\tanh^{-1}\left[e^{\phi_1}\sqrt{\frac{g_1+g_2}{g_1}}\right],\\
\phi&=&\ln\left[\frac{e^{-\phi_1}g_1+e^{\phi_1}(g_1+g_2)}{2g_1+g_2}\right],\\
A&=&\frac{1}{2}\phi_1-\ln(1-e^{4\phi_1})+\ln[(e^{2\phi_1}-1)g_1\nonumber \\
& &+e^{2\phi_1}g_2]+\frac{1}{2}\ln[g_1+(g_1+g_2)e^{2\phi_1}]
\end{eqnarray}
where the coordinate $\tilde{r}$ is related to $r$ by the relation $\frac{d\tilde{r}}{dr}=e^{\frac{\phi}{2}}$.
\\
\indent This solution preserves $N=4$ supersymmetry in three dimensions and describes an RG flow from $N=4$ SCFT in the UV with $SO(4)\times SO(3)$ symmetry to another $N=4$ SCFT in the IR with $SO(3)\times SO(3)$ symmetry at which the operator dual to $\phi_1$ is irrelevant. Although the number of supersymmetry is unchanged, the flavor symmetry $SO(3)$ in the UV is broken by the operator dual to $\phi_1$. We can also truncate out the vector multiplet scalars and study supersymmetric RG flows to non-conformal field theories as in the previous cases. However, we will not consider this truncation since it leads to similar structure as in the previous two gauge groups.
   
\section{Conclusions and discussions}\label{conclusion}
We have studied dyonic gaugings of $N=4$ supergravity coupled to six vector multiplets with compact and non-compact gauge groups $SO(4)\times SO(4)$, $SO(3,1)\times SO(3,1)$ and $SO(4)\times SO(3,1)$. We have identified a number of supersymmetric $N=4$ $AdS_4$ vacua within these gauged supergravities and studied several RG flows interpolating between these vacua. The solutions describe supersymmetric deformations of the dual $N=4$ SCFTs with different flavor symmetries in three dimensions. These deformations are driven by relevant operators of dimensions $\Delta=1,2$ which deform the UV SCFTs to other SCFTs or to non-conformal field theories in the IR.
\\
\indent For $SO(4)\times SO(4)$ gauge group, there are four supersymmetric $AdS_4$ vacua with $SO(4)\times SO(4)$, $SO(4)\times SO(3)$, $SO(3)\times SO(4)$ and $SO(4)$ symmetries. These vacua should correspond to $N=4$ conformal fixed points of $N=4$ CSM theories with $SO(4)$, $SO(3)$ and no flavor symmetries, respectively. We have found various RG flows interpolating between these critical points including RG flows connecting three critical points or a cascade of RG flows. These should be useful in holographic studies of $N=4$ CSM theories.  
\\
\indent In the case of non-compact $SO(3,1)\times SO(3,1)$ gauge group, we have found only one supersymmetric $AdS_4$ vacuum with $SO(3)\times SO(3)$ symmetry. We have given a number of RG flow solutions describing supersymmetric deformations of the dual $N=4$ SCFT to $N=4$ non-conformal field theories. The solutions with only scalar fields from the gravity multiplet non-vanishing give rise to physical singularities. Flows with vector multiplet scalars turned on, however, lead to physically unacceptable singularites. The mixed gauge group $SO(4)\times SO(3,1)$ also exhibits similar structure of vacua and RG flows with two supersymmetric $AdS_4$ critical points. 
\\
\indent Given our solutions, it is interesting to identify their higher dimensional origins in ten or eleven dimensions. Along this line, the result of \cite{S3_S3_reduction_Henning} and \cite{S3_S3_Dibitetto} on $S^3\times S^3$ compactifications could be a useful starting point for the $SO(4)\times SO(4)$ gauge group. The uplifted solutions would be desirable for a full holographic study of $N=4$ CSM theories. This should provide an analogue of the recent uplifts of the GPPZ flow describing a massive deformation of $N=4$ SYM \cite{GPPZ_Henning,GPPZ_Bobev}. The embedding of the non-compact gauge groups $SO(3,1)\times SO(3,1)$ and $SO(4)\times SO(3,1)$ would also be worth considering.    
\\
\indent Another direction is to find interpretations of the solutions given here in the dual $N=4$ CSM theories with different flavor symmetries similar to the recent study in \cite{5Dflow_bobev} for AdS$_5$/CFT$_4$ correspondence. The results found here is also in line with \cite{5Dflow_bobev}. In particular, scalars in the gravity multiplet are dual to relevant operators at all critical points. These operators are in the same multiplet as the energy-momentum tensor. Another result is the exclusion between the operators dual to the axion and vector multiplet scalars which cannot be turned on simultaneously as required by supersymmetry in the gravity solutions. It would be interesting to find an analogous result on the field theory side. 
\\
\indent A generalization of the present results to include more active scalars with smaller residual symmetries could provide more general holographic RG flow solutions in particular flows that break some amount of supersymmetry. We have indeed performed a partial analysis for $SO(3)_{\textrm{inv}}$ scalars. In this case, there are six singlets. It seems to be possible to have solutions that break supersymmetry from $N=4$ to $N=1$, but the scalar potential takes a highly complicated form. Therefore, we refrain from presenting it here. Solutions from other gauge groups more general than those considered here also deserve investigations. Finally, finding other types of solutions such as supersymmetric Janus and flows across dimensions to $AdS_2\times \Sigma_2$, with $\Sigma_2$ being a Riemann surface, would also be useful in the holographic study of defect SCFTs and black hole physics. Recent works along this line include \cite{orbifold_flow,warner_Janus,N3_Janus,Minwoo_4DN8_Janus,Guarino_AdS2_1,Guarino_AdS2_2,Trisasakian_AdS2}.   
\vspace{0.5cm}\\
{\large{\textbf{Acknowledgement}}} \\
This work is supported by The 90th Anniversary of Chulalongkorn University Fund (Ratchadaphiseksomphot Endowment Fund) and the Graduate School, Chulalongkorn University. P. K. is also supported by The Thailand Research Fund (TRF) under grant RSA5980037.
\appendix
\section{Useful formulae}
To convert an $SO(6)$ vector index $m$ to a pair of anti-symmetric $SU(4)$ fundamental indices $[ij]$, we use the following 't Hooft symbols
\begin{align}
&G_{1}^{ij} = \left[
\begin{array}{cccc}
 0 & i & 0 & 0 \\
 -i & 0 & 0 & 0 \\
 0 & 0 & 0 & -i \\
 0 & 0 & i & 0 \\
\end{array}
\right],
&G_{2}^{ij} = \left[
\begin{array}{cccc}
 0 & 0 & i & 0 \\
 0 & 0 & 0 & i \\
 -i & 0 & 0 & 0 \\
 0 & -i & 0 & 0 \\
\end{array}
\right],\notag\\
&G_{3}^{ij} = \left[
\begin{array}{cccc}
 0 & 0 & 0 & i \\
 0 & 0 & -i & 0 \\
 0 & i & 0 & 0 \\
 -i & 0 & 0 & 0 \\
\end{array}
\right],
&G_{4}^{ij} = \left[
\begin{array}{cccc}
 0 & -1 & 0 & 0 \\
 1 & 0 & 0 & 0 \\
 0 & 0 & 0 & -1 \\
 0 & 0 & 1 & 0 \\
\end{array}
\right],\notag\\
&G_{5}^{ij} = \left[
\begin{array}{cccc}
 0 & 0 & -1 & 0 \\
 0 & 0 & 0 & 1 \\
 1 & 0 & 0 & 0 \\
 0 & -1 & 0 & 0 \\
\end{array}
\right],
&G_{6}^{ij} = \left[
\begin{array}{cccc}
 0 & 0 & 0 & -1 \\
 0 & 0 & -1 & 0 \\
 0 & 1 & 0 & 0 \\
 1 & 0 & 0 & 0 \\
\end{array}
\right].
\end{align}
These matrices satisfy the relation
\begin{equation}
G_{mij}=-\frac{1}{2}\epsilon_{ijkl}G^{kl}_{m}=-(G_{m}^{ij})^{*}\, .
\end{equation}


\end{document}